\documentclass[%
 reprint,
superscriptaddress,
 amsmath,amssymb,
 aps,
prb,
longbibliography,
]{revtex4-1}
\usepackage{graphicx}
\usepackage{epstopdf}
\usepackage{dcolumn}
\usepackage{bm}
\usepackage{color}
\usepackage{hyperref}

\hypersetup{
    colorlinks=true,
    linkcolor=blue,
    filecolor=magenta,
    urlcolor=cyan,
}

\usepackage[caption=false]{subfig}
\usepackage{xr}
\usepackage{cleveref}
\usepackage{siunitx}

\newcommand{\beq}{\begin{equation}}
\newcommand{\eeq}{\end{equation}}
\newcommand{\bea}{\begin{eqnarray}}
\newcommand{\eea}{\end{eqnarray}}

\begin{document}

\preprint{}


\title{Coherent Control of Two-Dimensional Excitons}

\author{Christopher Rogers}
\email{cmrogers@stanford.edu}
\affiliation{%
 Ginzton Laboratory, Stanford University, 348 Via Pueblo, Stanford, CA 94305
}%
\author{Dodd Gray, Jr.}
\affiliation{%
 Ginzton Laboratory, Stanford University, 348 Via Pueblo, Stanford, CA 94305
}%
\author{Nathan Bogdanowicz}
\affiliation{%
 Ginzton Laboratory, Stanford University, 348 Via Pueblo, Stanford, CA 94305
}%
\author{Takashi Taniguchi}
\affiliation{%
National Institute for Materials Science, 1-1 Namiki, Tsukuba 305-0044, Japan
}%
\author{Kenji Watanabe}
\affiliation{%
National Institute for Materials Science, 1-1 Namiki, Tsukuba 305-0044, Japan
}%
\author{Hideo Mabuchi}
\email{hmabuchi@stanford.edu}
\affiliation{%
 Ginzton Laboratory, Stanford University, 348 Via Pueblo, Stanford, CA 94305
}%
%
%
%

\date{\today}

\pacs{Valid PACS appear here}
\maketitle


\textbf{
  Electric dipole radiation can be controlled by coherent optical feedback, as has previously been studied by modulating the photonic environment for point dipoles placed both in optical cavities \cite{Purcell_Effect, EnhancedAndInhibitedVisibleSpontaneousEMissionByAtoms, ControlledAtomicSpontaneousEmission} and near metal mirrors \cite{Drexhage1968, Drexhage1970}.
  In experiments involving fluorescent molecules \cite{Drexhage1968, Drexhage1970}, trapped ions \cite{Eschner2001_Atoms_Mirror, TrappedIonVacuumFieldLevelShifts} and quantum dots \cite{Stobbe2009_QD_Mirror} the point nature of the dipole, its sub-unity quantum efficiency, and decoherence rate conspire to severely limit any change in total linewidth.
  Here we show that the transverse coherence of exciton emission in the monolayer two-dimensional (2D) material $\mathbf{MoSe_2}$ removes many of the fundamental physical limitations present in previous experiments.
  The coherent interaction between excitons and a photonic mode localized between the $\mathbf{MoSe_2}$ and a nearby planar mirror depends interferometrically on mirror position, enabling full control over the radiative coupling rate from near-zero to 1.8 meV and a corresponding change in exciton total linewidth from 0.9 to 2.3 meV.
  The highly radiatively broadened exciton resonance (a ratio of up to $\mathbf{3:1}$ in our samples) necessary to observe this modulation is made possible by recent advances in 2D materials sample fabrication \cite{ExcitonicLinewidthApproachingHomogenousLimit, LargeExcitonicReflectivity, RealizationOfAnElectricallyTunableMirror}.
  Our method of mirror translation is free of any coupling to strain or DC electric field in the monolayer, which allows a fundamental study of this photonic effect.
  The weak coherent driving field in our experiments yields a mean excitation occupation number of $\mathbf{{\sim} 10^{-3}}$ such that our experiments correspond to probing radiative reaction in the regime of perturbative quantum electrodynamics \cite{RadiativePropertiesofAtomsNearAConductingPlane}.
  This system will serve as a testbed for exploring new excitonic physics \cite{ExcitonPolaritonCondensateReview} and quantum nonlinear optical effects \cite{Imamoglu_PRA_NonlinearMirror, Wild_PRL_QuantumNonlinearOptics}.
}

The transition metal dichalcogenides (TMDs) $\mathrm{MoSe_2}$ and $\mathrm{MoS_2}$ become direct band gap semiconductors when isolated in monolayer form \cite{AtomicallyThinMoS2, EmergingPhotoluminescence, MoSe2DirectBandgap}, transferring a significant fraction of the interband spectral weight to a strong and spectrally narrow excitonic resonance \cite{PRB2013_MoS2AbInitio, PRB2016_MoS2_abInitio}.
Coherence \cite{ExcitonValleyCoherence}, spin-valley interactions \cite{CoupledSpinValleyPhysics, ControlOfValleyPolarization}, strain effects \cite{BandgapEngineeringOfStrainedMoS2}, many-body electron physics \cite{ObservationOfBiexcitonsInMonolayerWSe2, FermiPolaronPolaritons, TightlyBoundTrions} and engineered confinement \cite{ElectricalControlofChargedCarriersInAtomicallyThin, LargeScaleQuantumEmitterArrays} have all been studied using TMD excitons.

Monolayer and few-layer TMDs were first prepared by mechanical exfoliation \cite{AtomicallyThinMoS2, EmergingPhotoluminescence, Exfoliation_1965} and were typically n-doped and inhomogeneously broadened by substrate roughness.
By adding electrostatic control via a gate, the semiconductor could be made neutral \cite{FermiPolaronPolaritons, TightlyBoundTrions}.
Encapsulation of TMDs in atomically flat hBN (hexagonal Boron Nitride) has enabled further improvements \cite{ExcitonicLinewidthApproachingHomogenousLimit, LargeExcitonicReflectivity, RealizationOfAnElectricallyTunableMirror}.
While some residual imperfections persist \cite{OpticalImagingOfStrain, ExploringAtomicDefectsInMoS2, UltrafastDynamicsofDefectAssistedRecombiniation, ScanningTunnelingMicroscopyOfGraphene}, sample qualities sufficient to manifest quantum coherent effects are now achievable.

Modifying the electromagnetic environment by using a mirror to engineer the local photonic density of states can affect the radiative decay rate of a dipole \cite{Purcell_Effect, Drexhage1970}.
In addition to those involving fluorescent molecules \cite{Drexhage1968, Drexhage1970}, trapped ions \cite{Eschner2001_Atoms_Mirror, TrappedIonVacuumFieldLevelShifts} and quantum dots \cite{Stobbe2009_QD_Mirror}, similar studies have been conducted with surface plasmon-polaritons \cite{PRL2018_SPP_Drexhage} and with an acoustic gong \cite{PRL_2016_Drexhage_Sound}.
For a perfect 0D dipole placed near a perfectly reflective spherical concave mirror, the radiative coupling and total linewidth could in principle be modulated from near zero to twice their vacuum values \cite{Drexhage1970}.
Experimentally, the modulation in total linewidth is much smaller due to the sub-unity quantum efficiency of real dipoles, their decoherence properties, and the use of planar mirrors (or finite numerical aperture \cite{Eschner2001_Atoms_Mirror}) for practical reasons, which partially obscure the interference effects.
For a planar mirror, interference effects on the total linewidth likewise decrease rapidly with mirror-dipole distance because of the high numerical aperture of the emission pattern.
However, the situation is different for excitons in 2D materials because the delocalized nature of the planar exciton leads to conservation of transverse momentum \cite{Farhan_PRB_2016_RadiativeLifetime, Daniel_MoS2_Nonlinearity}, meaning that the exciton emission is angularly restricted.
This opens the possibility of full manipulation of the radiative coupling even when the mirror is many wavelengths from the emitter.

The features that make this system a novel testbed for optical physics also make it  attractive for engineering applications.
Coupling mirror-membrane position to the frequency, linewidth and strength of a resonance is of interest for optomechanics.
For nonlinear and quantum optics, controllably reducing the intrinsic linewidth would greatly enhance nonlinearities \cite{Wild_PRL_QuantumNonlinearOptics}.
High quality TMDs grown by chemical vapor deposition \cite{ScalableSynthesisofUniformMonolayerMoS2, SynthesisOfLargeAreaMoS2} and then laser annealed to improve sample quality \cite{LaserAnnealing} offer a path towards scalable quantum engineering applications.

\begin{figure*}
    \centering
    \subfloat[ \label{fig:Sample:schematic}]
    {\includegraphics[width=0.37\textwidth]
    {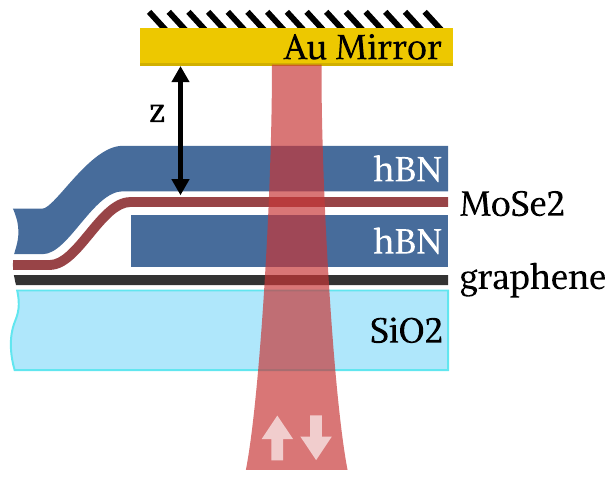}}
    \subfloat[ \label{fig:Sample:Image}]
    {\includegraphics[width=0.33\textwidth]
    {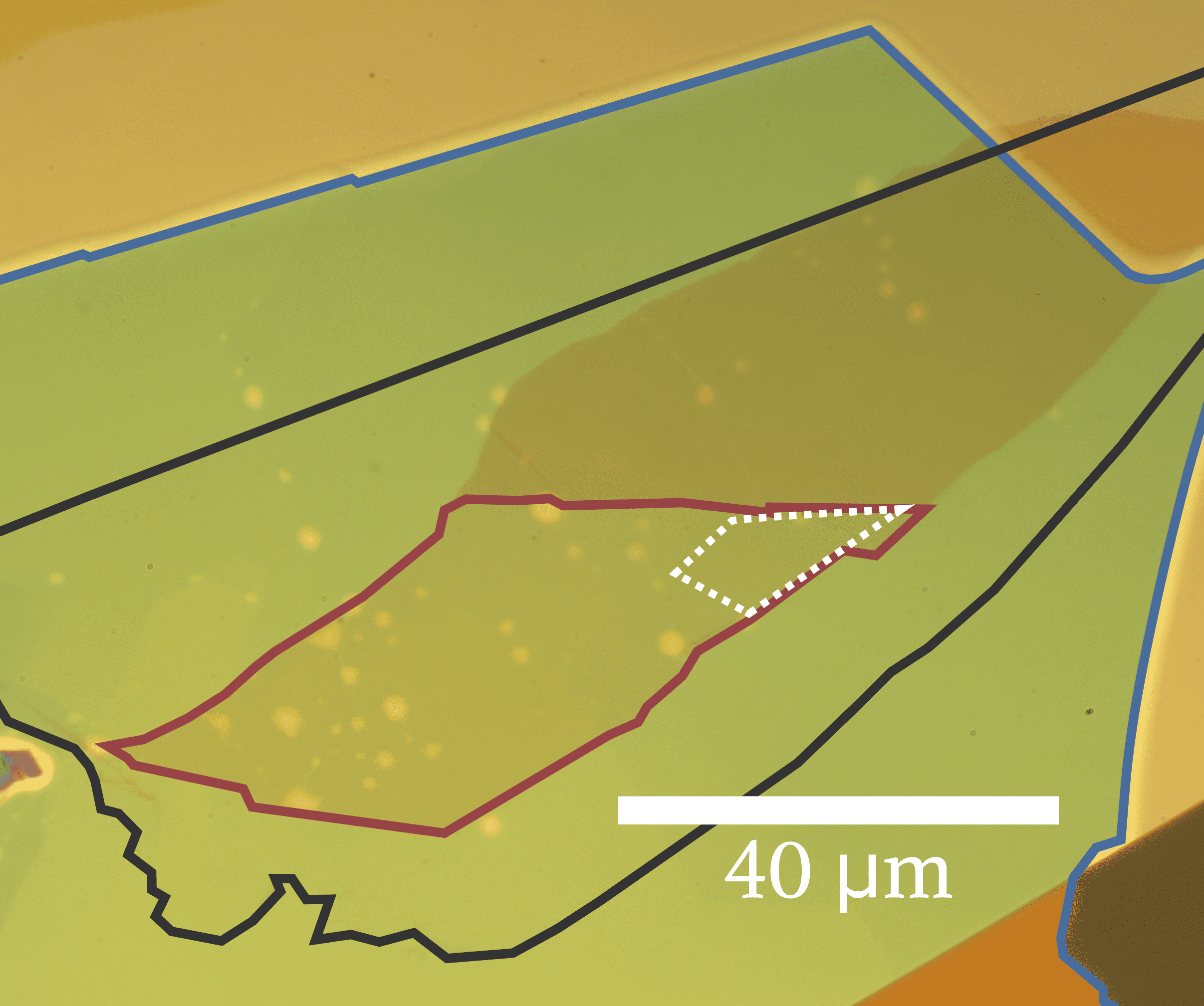}}

    \subfloat[ \label{fig:Sample:Maps}]
    {\includegraphics[width=0.44\textwidth]
    {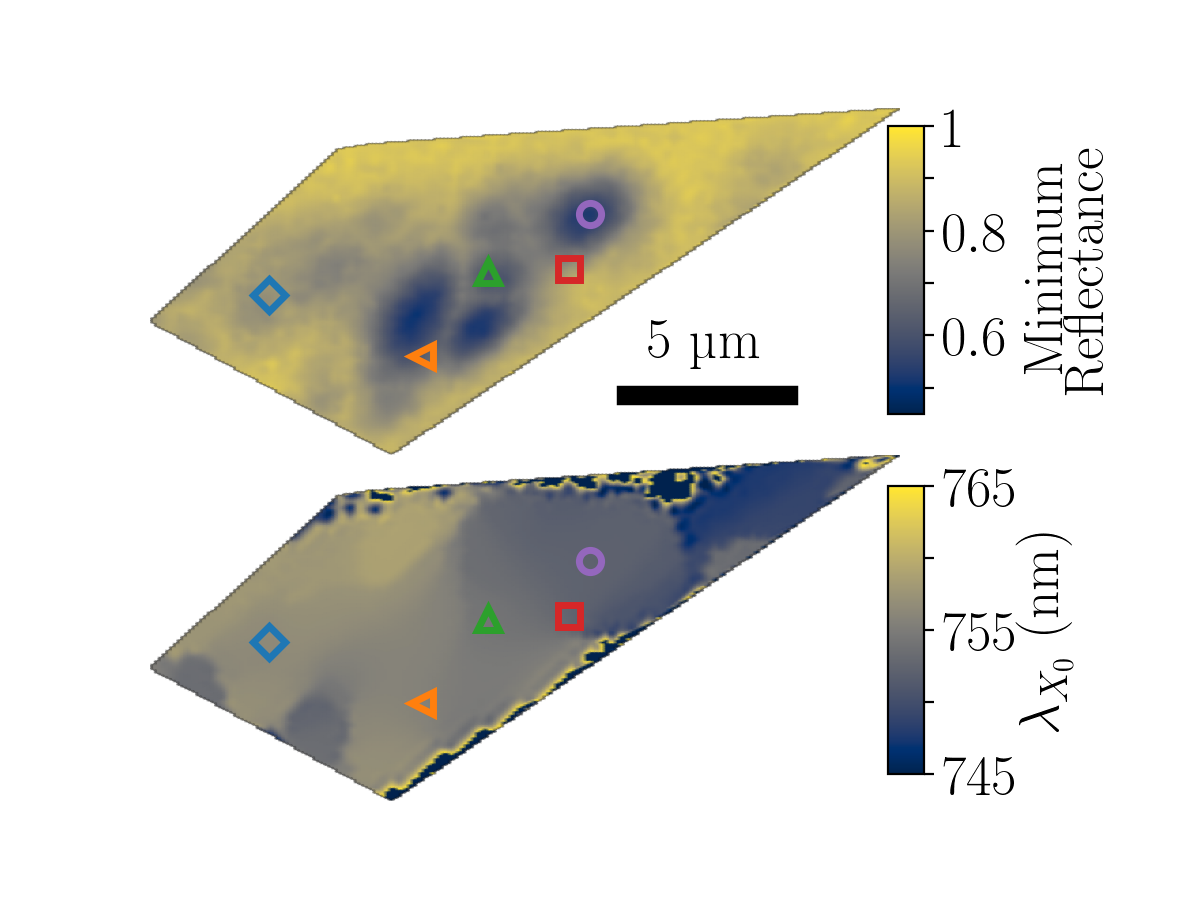}}
    \subfloat[ \label{fig:Sample:Spots}]
    {\includegraphics[width=0.3\textwidth]
    {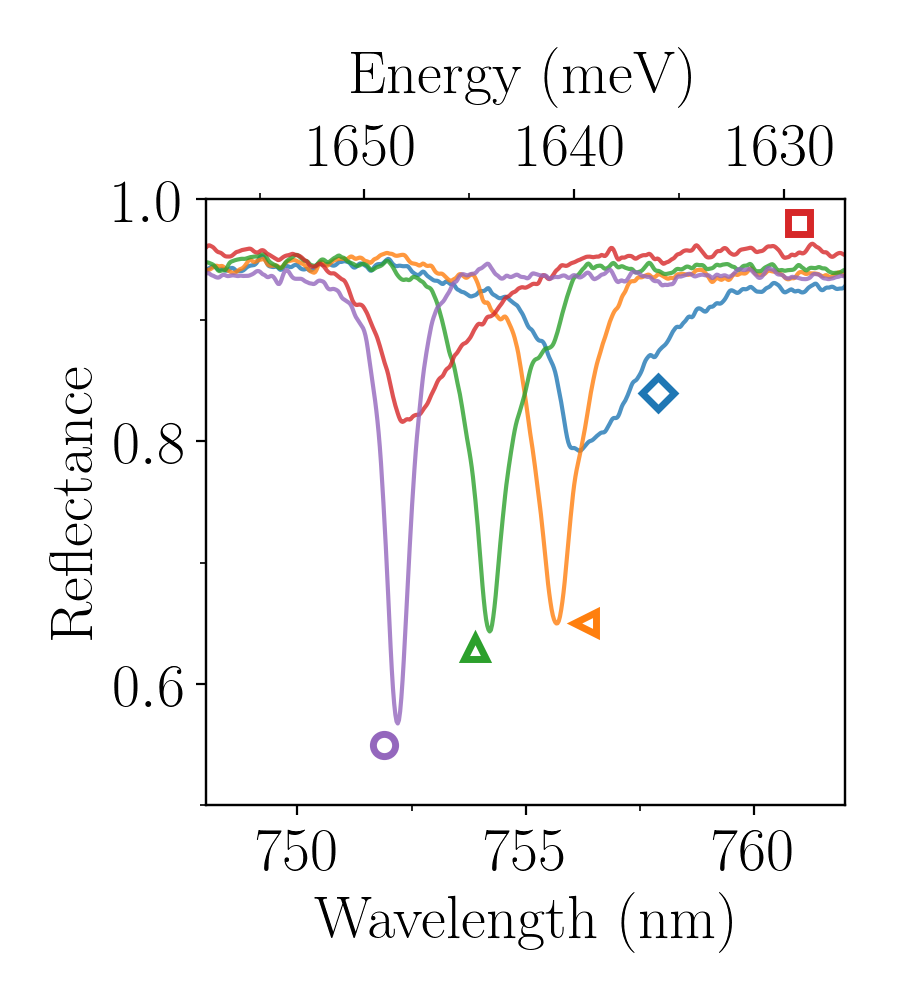}}
    \caption{\textbf{Sample Characterization.} (a) A schematic of the heterostructure device.
    The Au mirror is mounted on a mechanical actuator so that $z$ can be varied.
    (b) A microscope image of the sample.
    The monolayer $\mathrm{MoSe_2}$ region is outlined in burgundy.
    A few-layer graphene flake outlined in black on the bottom of the stack electrostatically isolates the $\mathrm{MoSe_2}$ from the $\mathrm{SiO_2}$ substrate.
    The bottom hBN is outlined in blue.
    (c) Spatial maps of the maximum dip in reflection at $X_0$, and its center wavelength $\lambda_{X_0}$ from a region of the sample corresponding to the area outlined with dashed white lines in (b).
    (d) Selected reflection spectra corresponding to the marked positions in (c).
    These spectra were collected at 4 K with the mirror positioned slightly away from maximum destructive interference.
    \label{fig:Sample}}
\end{figure*}

We report the effect of varying the distance between the monolayer semiconductor $\mathrm{MoSe_2}$ and a metal mirror on the $\mathrm{MoSe_2}$ exciton resonance ($X_0$).
A low-finesse photonic mode is formed between the mirror and $\mathrm{MoSe_2}$, and light coupling out of this cavity interferes with light directly emitted by the exciton.
As the mirror is translated, the interference condition at the $\mathrm{MoSe_2}$ varies between  destructive and constructive, strongly modifying the reflection of the device.
The magnitude of the reflection at $X_0$ can vary from near zero to near unity, and the absorption varies in a complementary way.
This interference condition also affects the radiative coupling of $X_0$ to the environment, and
at maximal destructive interference the coupling can be almost entirely suppressed, in theory limited only by mirror losses.
Conversely, at maximal constructive interference this coupling is twice its vacuum value.
Since $X_0$ is primarily radiatively broadened, this modulation of the radiative coupling induces a similar effect on the total linewidth.

In the experiment we fabricate heterostructures of $\mathrm{MoSe_2}$ encapsulated in hBN, and then transfer these stacks onto fused silica substrates  \cite{FastPickupTechnique, HotPickupTechnique}.
A microscope image of the sample used for the data presented in this paper is shown in Fig. \ref{fig:Sample:Image}, and a schematic of the experiment is shown in Fig. \ref{fig:Sample:schematic}.
Experiments are conducted within an optical cryostat at 4 K.
A gold mirror on a mechanical actuator is placed in close proximity to the $\mathrm{MoSe_2}$ heterostructure.
The mirror is translated along the optical axis, and at selected $z$ positions reflectance measurements are made using a grating spectrometer.
Note that $z$ is the {\em optical} path length between the mirror and the $\mathrm{MoSe_2}$.
This method of mirror translation isolates the effect of coherent electromagnetic feedback since it is entirely free of coupling to strain or electric field in the TMD.
More detail on the samples and the experiment can be found in the methods sections \ref{subsec:Methods:Samples} and \ref{subsec:Methods:Setup}.
Because we excite with ${\sim} 15$ nW of continuous-wave optical power with a bandwidth of 300 nm the photon rate at the sample is ${\sim}60$ GHz, and considering only optical power resonant with $X_0$ the rate is ${\sim}0.4$ GHz.
Taking into account the exciton decay rate of ${\sim}2$ meV $\approx 480$ GHz, the excitation occupation number during the measurement is very low, ${\sim} 10^{-3}$.

Maps of the magnitude of the dip in reflectance at $X_0$ and its center frequency/wavelength ($\omega_{X_0}$, $\lambda_{X_0}$) are shown in Fig. \ref{fig:Sample:Maps}.
The mirror is near but not at maximum destructive interference.
As observed by others \cite{LargeExcitonicReflectivity, ExcitonicLinewidthApproachingHomogenousLimit}, there is inhomogeneity on a few-micron scale in both $\lambda_{X_0}$ and the magnitude of the reflection.
Nonetheless, some areas of the sample are radiatively broadened.
In Fig. \ref{fig:Sample:Spots}, spots are selected to show both a range of sample quality and $\lambda_{X_0}$.

\begin{figure*}
    \centering
    \subfloat[ \label{fig:Data:Heatmap}]
    {\includegraphics[height=0.34\textwidth]
    {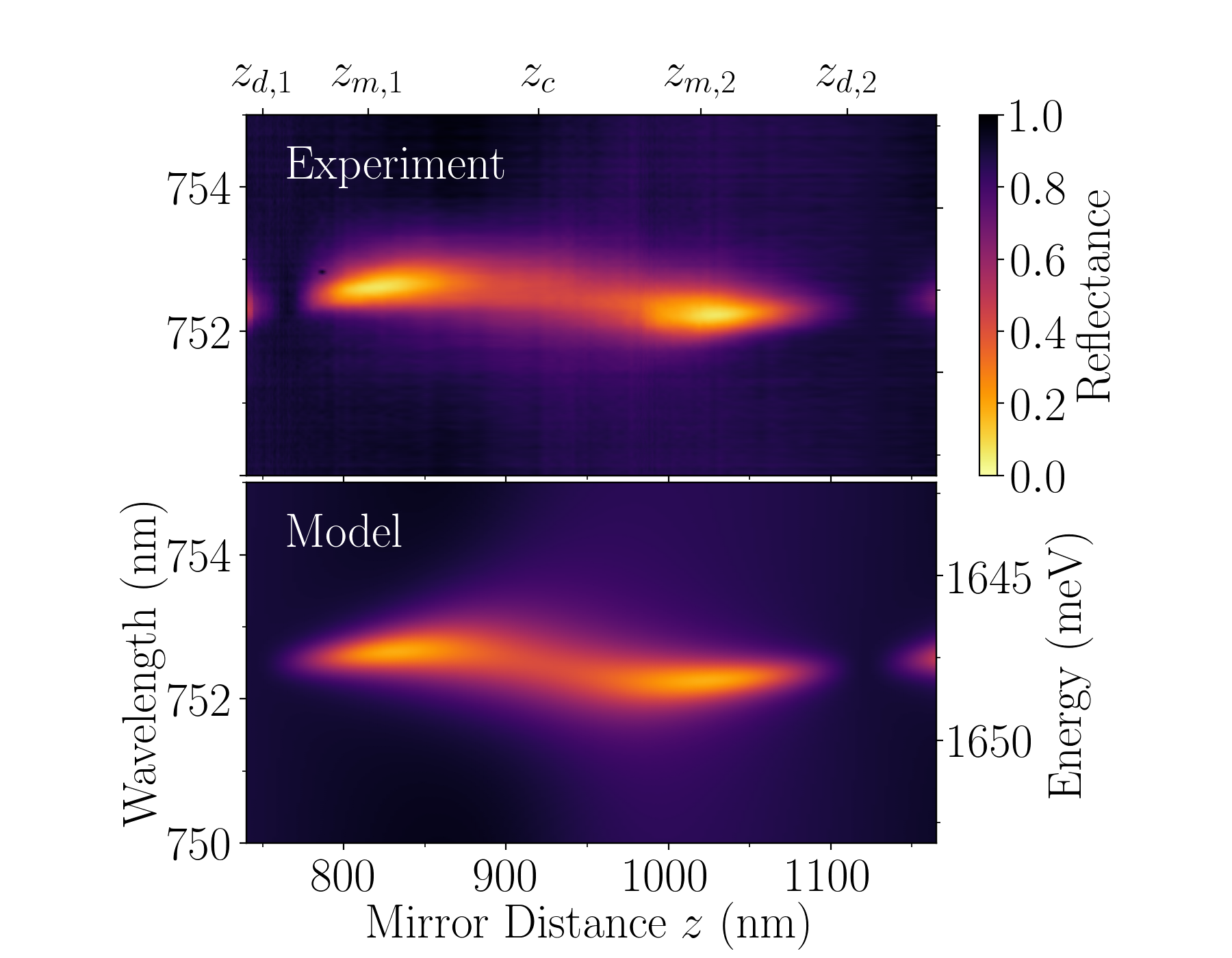}}
    \subfloat[ \label{fig:Data:InterferenceSchematic}]
    {\includegraphics[height=0.34\textwidth]
    {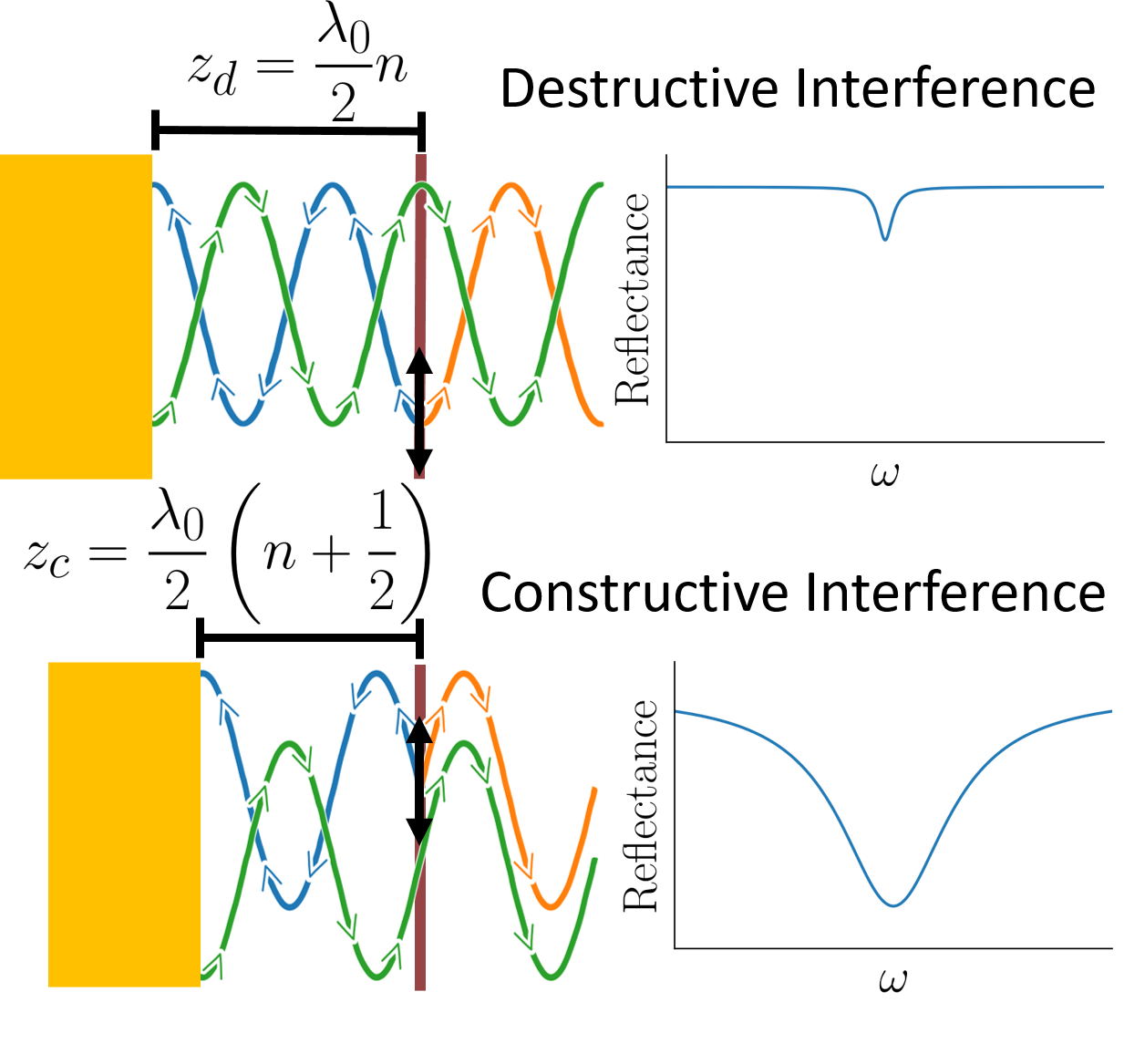}}

    \subfloat[ \label{fig:Data:LineCuts}]
    {\includegraphics[height=0.39\textwidth]
    {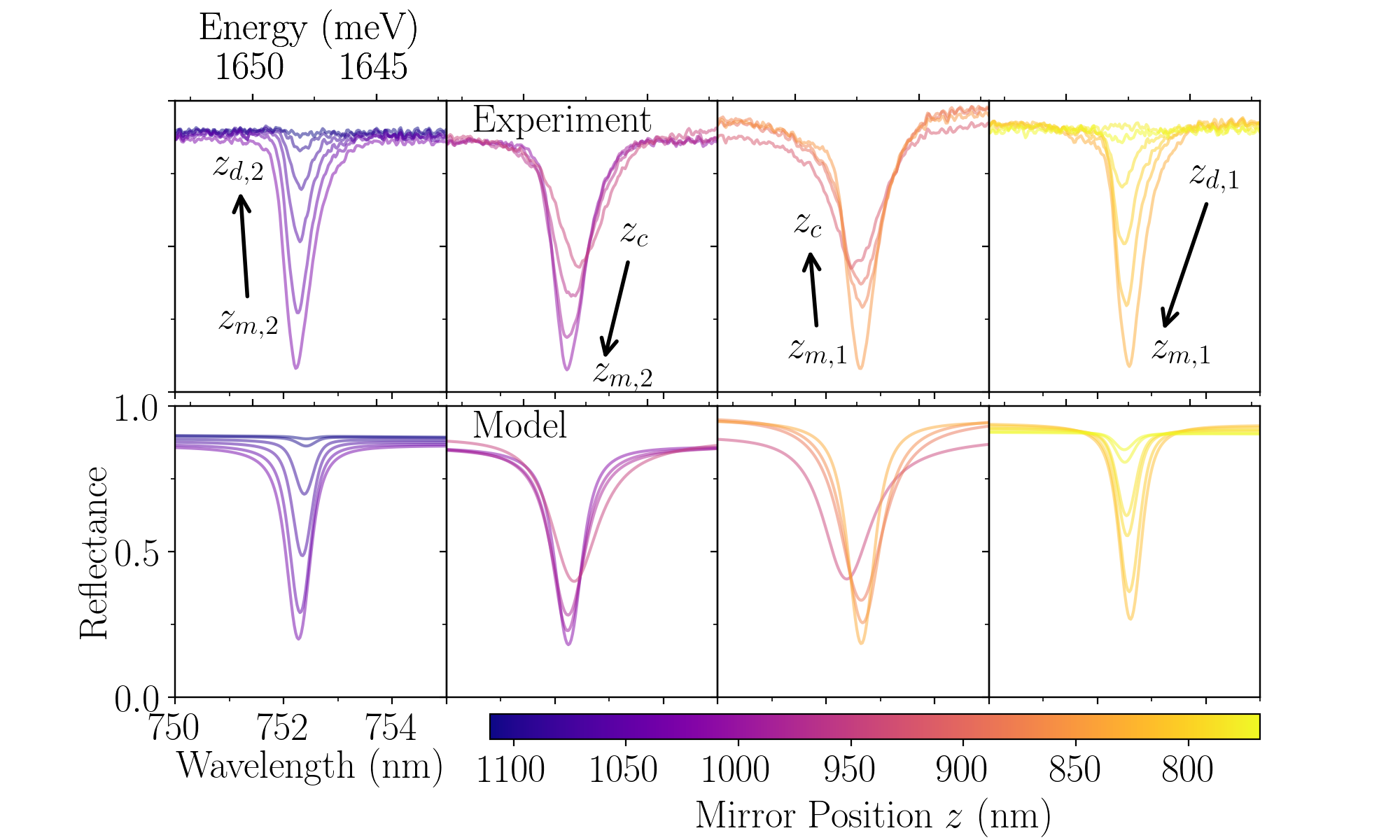}}
    \subfloat[ \label{fig:Data:MaxMinWidth}]
    {\includegraphics[height=0.39\textwidth]
    {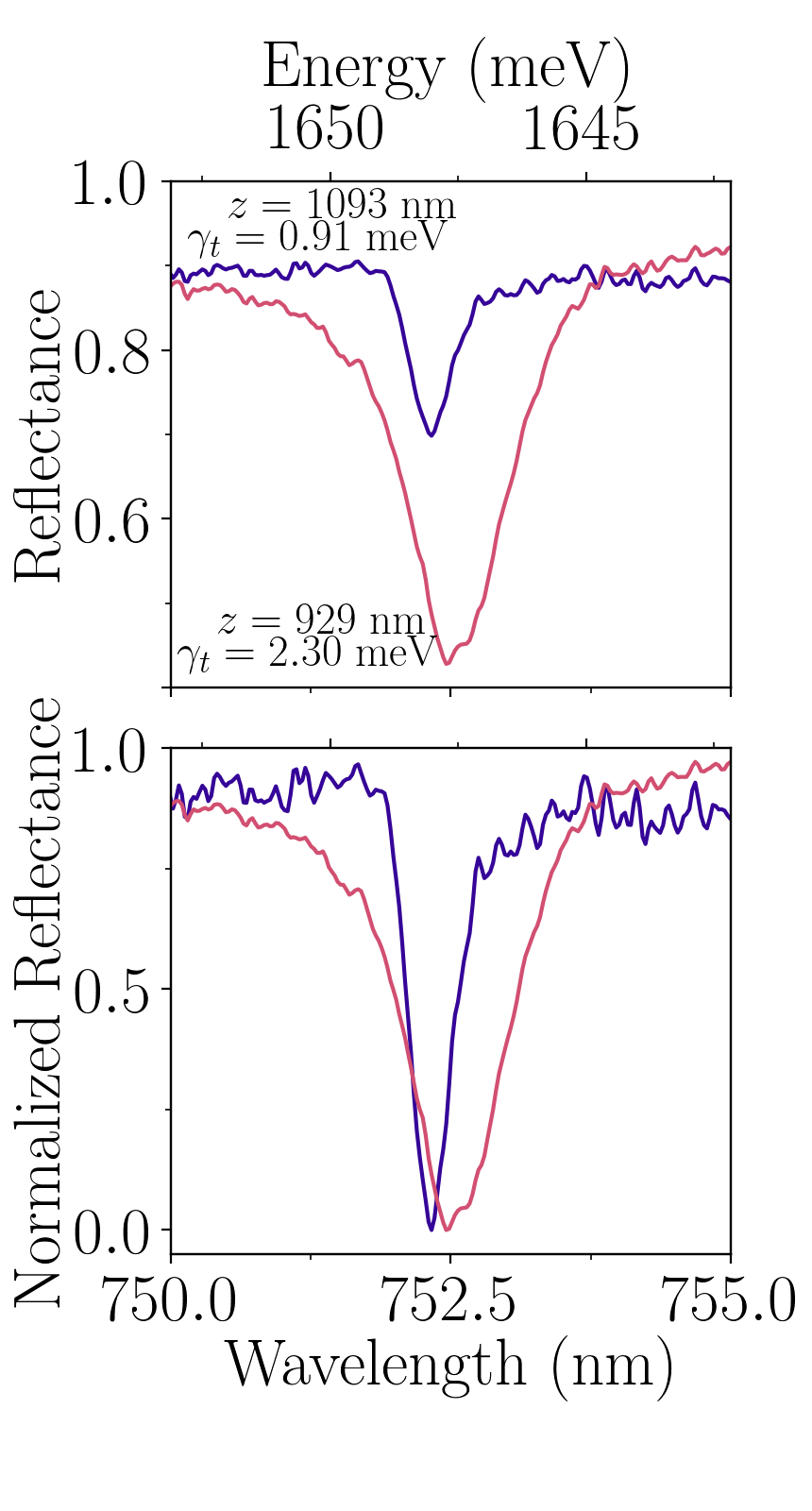}}
    \caption{\textbf{Experimental and Modeled Reflectance.} (a) Measured and modeled reflectance spectra near the $X_0$ resonance as $z$ is varied over a full fringe.
    Measurements at 4 K.
    Note that this data comes from near the region marked with a purple circle in Fig.  \ref{fig:Sample:Maps}.
    (b) A schematic representation of the effect of mirror position $z$ on the the exciton resonance.
    The black double-headed arrow represents the exciton electric dipole, and the blue and orange curves represent the electric field emitted towards and away from the mirror, respectively.
    The electric field reflected by the mirror is represented in green.
    To the right of the monolayer the fields interfere either constructively or destructively depending on mirror position.
    The corresponding schematic plots represent the modulation of amplitude and linewidth of the $X_0$ feature.
    The $z_d$ and $z_c$ values shown respectively for the destructive and constructive interference cases respectively, assume a perfect mirror with zero skin-depth.
    For simplicity, multiple reflections and the full heterostructure have been omitted.
    (c) Selected line cuts of the measured and modeled reflectance in the spectral region of $X_0$.
    The black arrows indicate increasing $z$.
    (d) Measured reflectance, both absolute and normalized, at two $z$ positions highlighting the modulation of total linewidth.
    \label{fig:Data}}
\end{figure*}

\begin{figure*}
    \centering
    \subfloat[ \label{fig:Parameters:ExtractedLinewidth}]
    {\includegraphics[width=0.287\textwidth]
    {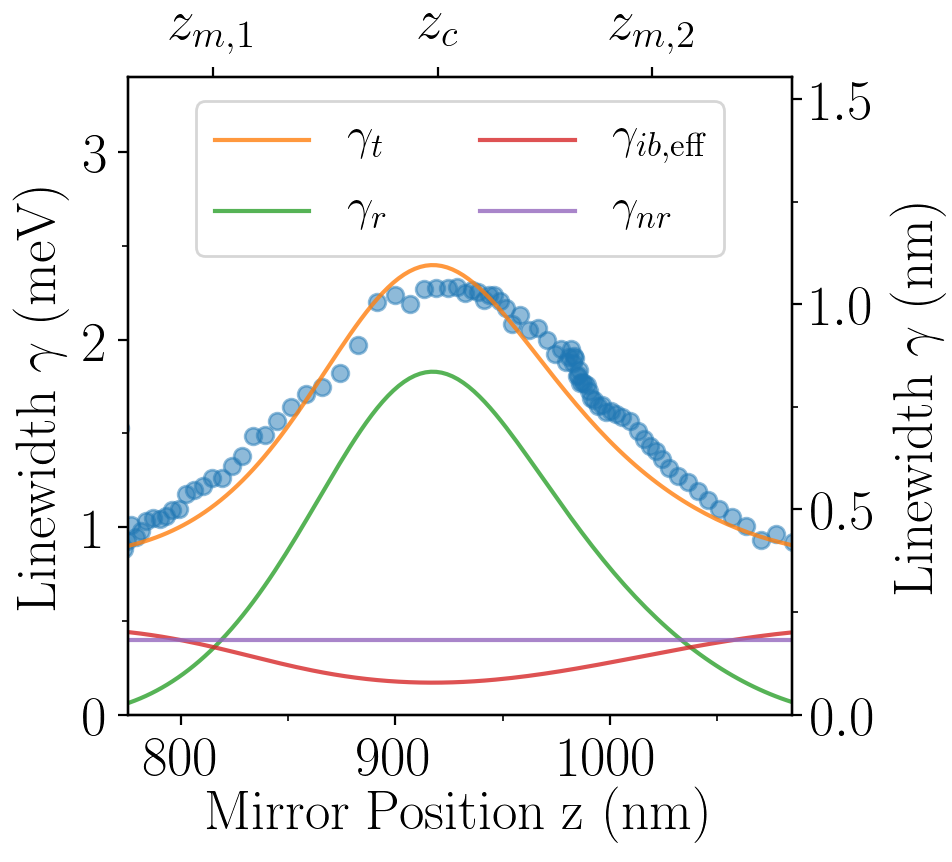}}
    \subfloat[ \label{fig:Parameters:ExtractedPosition}]
    {\includegraphics[width=0.345\textwidth]
    {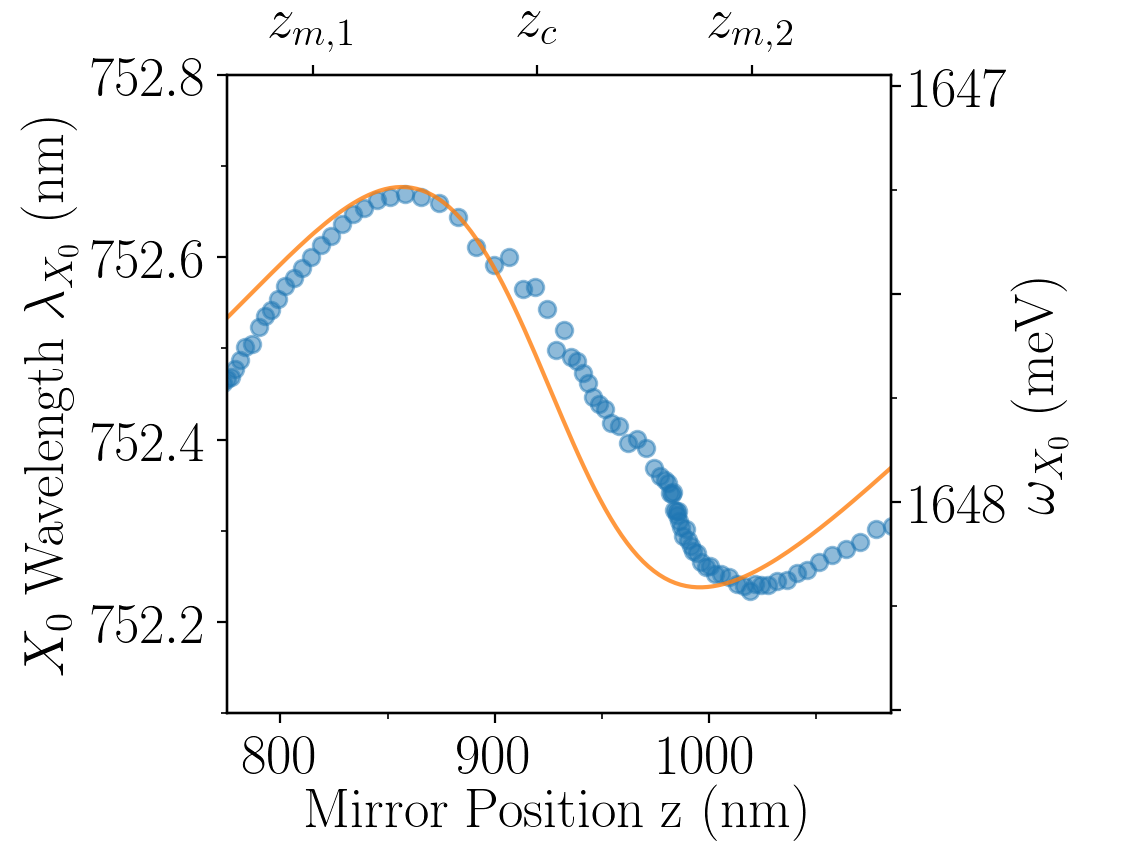}}
    \subfloat[ \label{fig:Parameters:ExtractedDip}]
    {\includegraphics[width=0.265\textwidth]
    {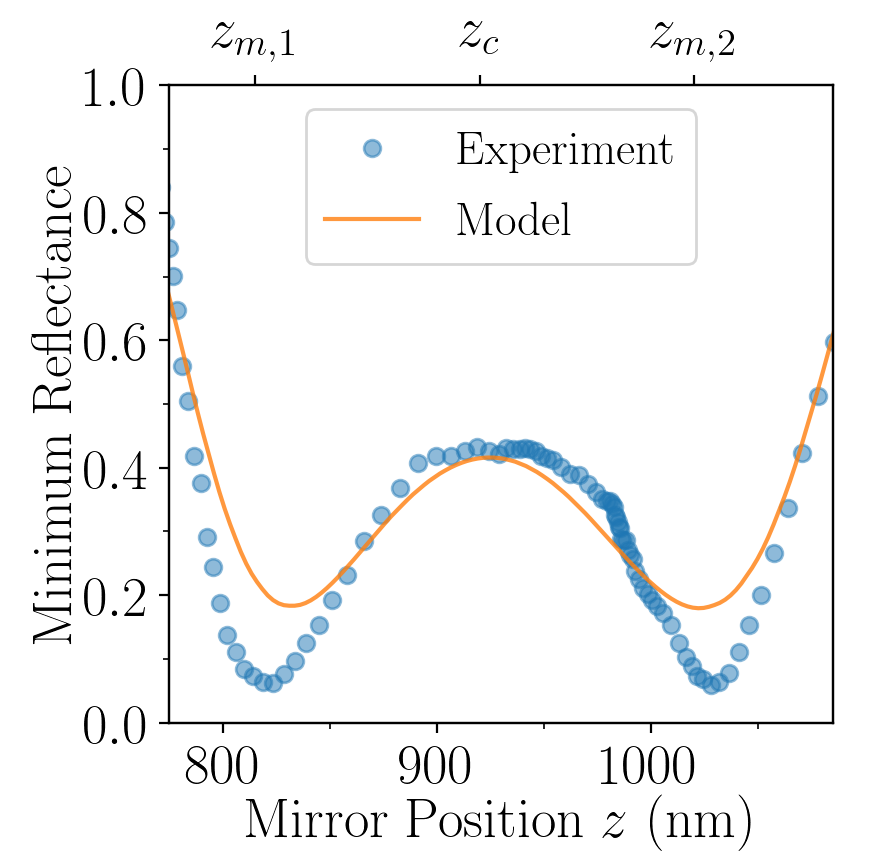}}

    \subfloat[ \label{fig:Parameters:Drexhage}]
    {\includegraphics[width=0.5\textwidth]
    {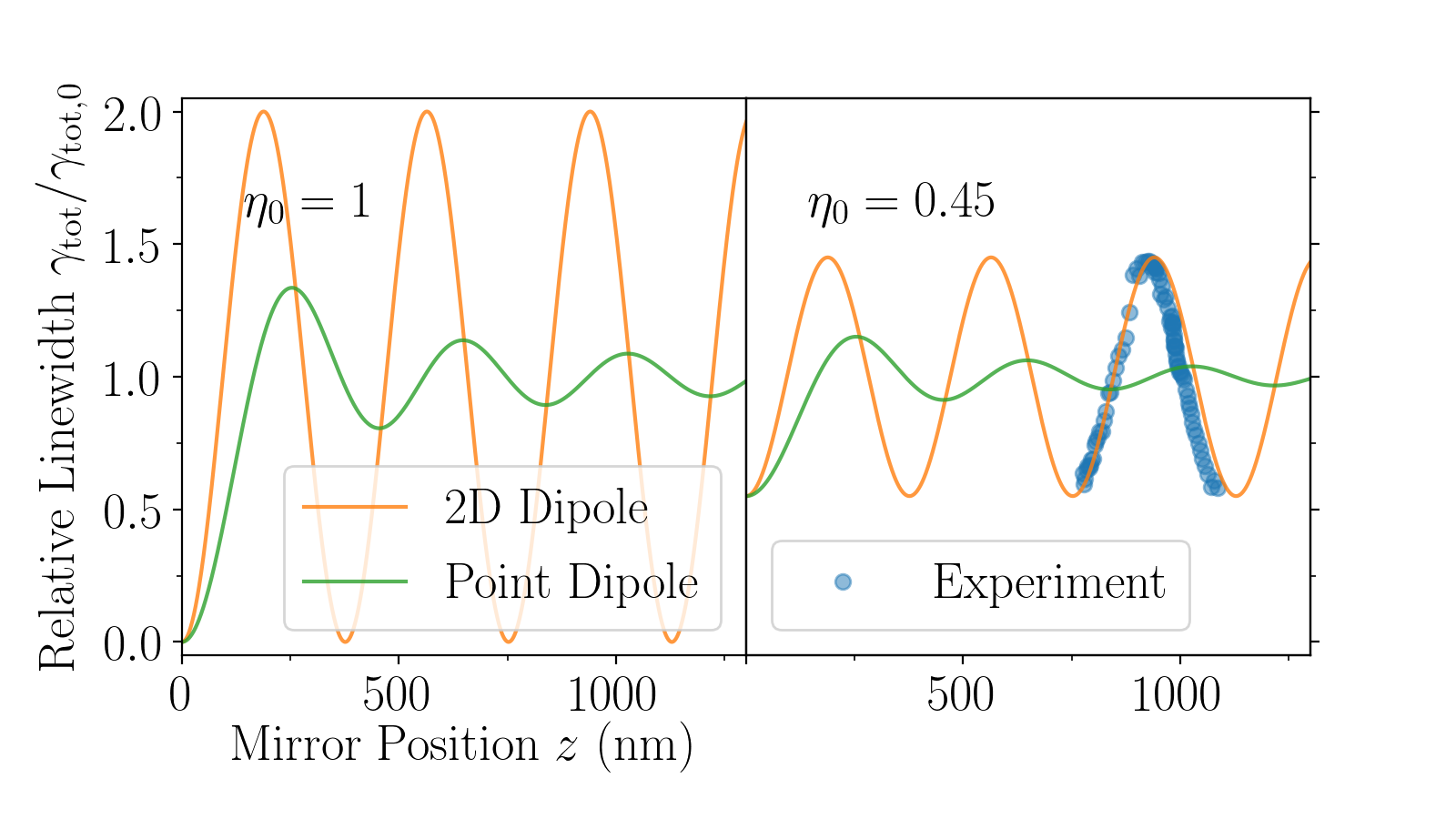}}
    \caption{\textbf{Extracted and Modeled Linewidths.} (a) The FWHM linewidth $\gamma_\mathrm{tot}$ both from the model and extracted from the experimental data.
    Note that we cannot extract linewidth data over the full range of the experimental data, since near $z_d$ the $X_0$ resonance is almost completely extinguished.
    Also shown are $\gamma_{r}$, $\gamma_{nr}$ and $\gamma_{ib, \mathrm{eff}}$ from the model.
    (b) Center frequency $\omega_{X_0}$ for both model and experiment.
    (c) Minimum reflectance for both model and experiment.
    (d) Simplified models of the total linewidth modulation for both a point and 2D dipole, assuming a perfect mirror with zero skin depth.
    On the left is shown the ideal case with coherent quantum efficiency in vacuum $\eta_0 = 1$, and on the right with $\eta_0 = 0.45$ alongside the experimental data.
    \label{fig:Parameters}}
\end{figure*}

A heat map of the reflectance as a function of mirror position is shown in Fig. \ref{fig:Data:Heatmap}, along with selected line cuts of the same data  in Fig. \ref{fig:Data:LineCuts}.
The $X_0$ resonance appears as a dip (the central bright band) that varies in magnitude, width and center frequency as the mirror is translated across a full fringe.
We define $z_{d}$ and $z_c$ as the mirror positions for maximal destructive and constructive interference respectively, as in Fig. \ref{fig:Data:InterferenceSchematic}.
When the reflection from the mirror interferes destructively with that from the $\mathrm{MoSe_2}$, the radiative coupling of $X_0$ becomes very small and the dip disappears below the noise floor ($z_{d,1}$ and $z_{d,2}$ in Fig. \ref{fig:Data:Heatmap}).
Surprisingly, the minimum reflection over $z$ does not occur at $z_c$, but rather occurs at each of two mirror positions on either side of $z_c$.
At these two reflection minima ($z_{m,1} = 815$ nm and $z_{m,2} = 1020$ nm) the reflectance is ${\sim}8 \%$, while in between it reaches $45\%$ at $z_c$.

This surprising effect is due to an interplay between the radiative coupling rate ($\gamma_r$) and non-radiative coupling rate ($\gamma_{nr}$).
At $z_c$, the exciton is primarily radiatively broadened and $\gamma_r$ is larger than $\gamma_{nr} + \gamma_{ib, \mathrm{eff}}$, where $\gamma_{ib, \mathrm{eff}}$ is the effective contribution to the total linewidth from a Gaussian inhomogeneous broadening $\gamma_{ib}$.
We define $\gamma_\mathrm{tot} = \gamma_r + \gamma_{nr} + \gamma_{ib, \mathrm{eff}}$ as the total linewidth.
In an ideal material ($\gamma_r \gg \gamma_{nr}, \gamma_{ib, \mathrm{eff}}$) the reflectance would approach unity here.
However, in our real sample $\gamma_{nr}$ and $\gamma_{ib, \mathrm{eff}}$ reduce the reflectance feature at $z_c$ to $45\%$.
When the mirror is moved in either direction from $z_c$, $\gamma_r$ and $\gamma_r/\gamma_{nr}$ are reduced so that the exciton begins to absorb more light, causing the reflectance feature to deepen and eventually reach its minimum value at $z_{m,1/2}$.
It also begins to spectrally narrow, since the contribution of $\gamma_r$ to the total linewidth $\gamma_\mathrm{tot}$ is reduced.
As the mirror is moved even farther away from $z_c$ past $z_{m,1/2}$, $\gamma_r$ continues to decrease and the $X_0$ feature shrinks while spectrally narrowing until it eventually disappears.

Note that the center frequency $\omega_{X_0}$ of the dip changes with mirror position as well.
When the light reflected from the mirror is exactly in- or out-of-phase with that back-emitted from the $\mathrm{MoSe_2}$, the dip is at its `vacuum' frequency ($\omega_0$), and $\omega_{X_0} = \omega_0$.
However, away from either of these positions dispersion over the $X_0$ resonance causes a spectrally asymmetric interference condition, which shifts the effective line center $\omega_{X_0}$.

Plots of the modeled reflectance are shown in Figs. \ref{fig:Data:Heatmap} and \ref{fig:Data:LineCuts}.
We model the TMD exciton using a Lorentzian susceptibility, shown in Eq. \ref{eq:exciton_chi}, which accounts for radiative broadening $\gamma_r$ and non-radiative broadening $\gamma_{nr}$ \cite{LargeExcitonicReflectivity, WootenOpticalPropertiesOfSolids}.
The reflectance is calculated for the full heterostructure and mirror using a transfer matrix method \cite{TranferMatrixMethod} including the effect of inhomogeneous broadening $\gamma_{ib}$.
Selected line cuts were simultaneously fit to the model to find global values for $\gamma_r$, $\gamma_{nr}$, $\gamma_{ib}$ and $\omega_{X_0}$.
See the methods section \ref{subsec:Methods:Model} for more details.
Both qualitatively and quantitatively, the model matches closely with the experimental data.
One slight difference is that the deepest reflectance feature obtained over $z$ is smaller in the experiment ($8 \%$) than the model ($18 \%$), which is likely due to a combination of pure dephasing (which is not included in the model), spectrally structured inhomogeneous broadening, mode-mismatch between the interfering reflected beams, and diffraction effects.
We attribute the slight discrepancy (distortion along $z$) between the model and experiment on the left in Fig. \ref{fig:Data:Heatmap} to small imperfections in the $z$ calibration of the data.

From both the experimental data and the model we extract full-width-half-max (FWHM) linewidths, which are shown in Fig. \ref{fig:Parameters:ExtractedLinewidth}.
As a function of mirror position, the linewidth $\gamma_\mathrm{tot}$ varies from ${\sim}0.9$ meV near $z_d$ to ${\sim}2.3$ meV at $z_c$ for a total modulation of ${\sim}2.5 \times$, while in the model it varies from 0.9 to 2.4 meV, or ${\sim}2.6 \times$.
This modulation can also be clearly seen in Fig. \ref{fig:Data:MaxMinWidth}.
The change in $\gamma_r$ of the $X_0$ resonance is the primary cause of the change in $\gamma_\mathrm{tot}$, and the values of $\gamma_r$ extracted from the model vary from near zero at  $z_d$ to 1.8 meV (440 GHz) at  $z_c$.
Near $z_d$, $\gamma_\mathrm{tot}$ is dominated by the contribution of $\gamma_{nr}$ and $\gamma_{ib, \mathrm{eff}}$ while at $z_c$ radiative coupling dominates, with a ratio of $\gamma_r/(\gamma_{nr} + \gamma_{ib, \mathrm{eff}}) \sim 3$.
Modeled and experimental values for $\omega_{X_0}$ and the minimum reflectance shown in Figs. \ref{fig:Parameters:ExtractedPosition} and \ref{fig:Parameters:ExtractedDip} agree as well.
The line shift of ${\sim}1$ meV (240 GHz) is significant relative to $\gamma_\mathrm{tot}$.

Lastly, in Fig. \ref{fig:Parameters:Drexhage} we compare our data to a simplified model of both a 2D dipole and a point dipole.
The 2D case highlights that the transverse coherence and delocalized nature of the exciton causes light emission into specific modes rather than the full numerical aperture.
We define the coherent quantum efficiency in vacuum $\eta_0 = \gamma_r/\gamma_\mathrm{tot}$, which differs from the incoherent quantum efficiency $\gamma_r/(\gamma_r +\gamma_{nr})$.
For our purpose, coherent quantum efficiency is the relevant quantity because the linewidth modulation effect depends on the coherent interference of emitted waves.
Using models described in methods section \ref{subsec:Methods:Linewidths}, we show that for an ideal 2D dipole ($\eta_0 =1$) and a perfect planar mirror, the linewidth can be fully modulated even when the mirror is far from the dipole.
This is not true in the 0D dipole case, because integrating emission over the full numerical aperture obscures the interference effect as $z$ increases.
Also shown is a plot for $\eta_0=0.45$ chosen to match the superimposed experimental data.
Note that the peak ratio $ \gamma_r/(\gamma_{nr} + \gamma_{ib, \mathrm{eff}}) \sim 3$ extracted from the reflectance model does not match that of $ \gamma_r/(\gamma_{nr} + \gamma_{ib, \mathrm{eff}}) \sim 1.8$  we expect for $\eta_0=0.45$ from this simplified model.
This discrepancy is primarily due to $\gamma_{ib, \mathrm{eff}}$ varying with $z$ so as to partially counteract the change in $\gamma_r$, as seen in Fig. \ref{fig:Parameters:ExtractedLinewidth}.
Because the intrinsically Lorentzian exciton feature is convolved with a Gaussian inhomogeneous broadening of width $\gamma_{ib}$ to form a Voight profile, the effective inhomogeneous broadening $\gamma_{ib, \mathrm{eff}}$ is larger when the total linewidth $\gamma_\mathrm{tot}$ is small \cite{EmpiricalApproximationVoight, VoightBriefReview}.
Regardless, the amplitude and phase of the experimental $\gamma_\mathrm{tot}$ values highlight that no 0D dipole in front of a flat mirror (even with a perfect mirror and $\eta_0=1$) could produce the behavior seen in the experiments.

We have demonstrated coherent control over an exciton resonance by positioning a mirror in close proximity to the monolayer semiconductor $\mathrm{MoSe_2}$, showing near complete modulation of the reflection at $X_0$.
The concurrent change in radiative coupling rate induces a change in total linewidth of ${\sim}2.4\times$, demonstrating the dominant role of radiative coupling for excitons in monolayer $\mathrm{MoSe_2}$ and serving as an important verification of theoretical models used to describe excitonic physics in TMD materials.
For engineering applications, the modulation of the $X_0$ resonance could greatly enhance optomechanical couplings, while the effective enhancement of nonlinearities \cite{Wild_PRL_QuantumNonlinearOptics} is useful for nonlinear optics and quantum optics.
Our strain free and DC electric field free method of mirror positioning has allowed us to study the underlying photonic interference effect present in the system, and the unprecedented control over the radiative coupling of an excitonic resonance paves the way for many future experiments.
\newline\newline
{\em Note:} During preparation of this manuscript we became aware of a preprint presenting similar work by You Zhou, et. al. \cite{ControllingExcitonsWithAMirror}.

\section{\label{sec:Methods} Methods}
\subsection{\label{subsec:Methods:Samples} Sample Preparation}
We fabricate heterostructures like the one shown in Figs. \ref{fig:Sample:schematic} and \ref{fig:Sample:Image} using a dry pickup transfer technique \cite{FastPickupTechnique, HotPickupTechnique}.
First, hBN, $\mathrm{MoSe_2}$ and graphite are mechanically exfoliated \cite{AtomicallyThinMoS2, EmergingPhotoluminescence, Exfoliation_1965} onto 300 nm $\mathrm{SiO_2}$ on Si substrates.
The substrates are then observed under an optical microscope to identify suitable flakes.
Polycarbonate (PC) `stamps' are made by affixing a thin PC film to a piece of polydimethylsiloxane (PDMS) on a glass slide.
This stamp is then used to sequentially pick up the mechanically exfoliated flakes by bringing the stamp slowly into contact with a flake on the exfoliation substrate.
In our case, we first pick up the `top' hBN, then the monolayer $\mathrm{MoSe_2}$, then the `bottom' hBN and finally the few-layer graphene flake.
This stack (including the PC film) is transferred to a glass substrate.
The PC is removed by dissolution in chloroform.
Note that the $\mathrm{MoSe_2}$ protrudes from the bottom hBN so that it contacts the few-layer graphene flake, shorting the two to each other.
This helps to electrostatically isolate the $\mathrm{MoSe_2}$ from the substrate.
More details on sample preparation are given in the supplemental material.

The mirrors are prepared by taking a small glass substrate ($\sim 1$ mm x 1 mm in lateral dimensions) and affixing it to a larger carrier substrate.
This is then coated in 120 nm of gold, with a 3 nm titanium adhesion layer.

\subsection{\label{subsec:Methods:Setup} Experimental Setup}
The experiment is conducted in an optical cryostat (Montana Instruments Nanoscale Workstation).
All measurements were conducted at a nominal temperature of 4 K and a pressure of $1\cdot 10^{-7}$ Torr.
The sample is attached to a fixed mount.
A gold mirror is mounted on a slip-stick piezo mirror mount (Janssen Precision Engineering) and brought close to the sample.
The slip-stick piezo mirror mount is used to translate the mirror relative to the sample.
A microscope objective ($20\times$, 0.4 numerical aperture, Olympus MSPLAN) inside the cryostat was used to focus the beam onto the sample.

Light from either a lamp (Thorlabs SLS201) or a supercontinuum laser (NKT Photonics SuperK) is focused into a single mode optical fiber.
This light is coupled into a home-built confocal microscope setup, which focuses the beam onto the sample.
The collected reflected light is then sent to a home-built grating spectrometer for measurement.
In order to translate the focus of the beam in three dimensions at the sample, the confocal microscope has a double 4f system.
One tip-tilt mirror in the second 4f section translates the beam in the transverse plane, while a lens in the first 4f system is translated along the optical axis to move the beam focus along the optical axis.
More details of the experimental setup can be found in the supplemental information.
Each spectrum is normalized to a spectrum taken at a flake-free area on the substrate.
Measurements were automated using the python instrument control package Instrumental \cite{Instrumental}, available on GitHub at \url{https://github.com/mabuchilab/Instrumental}.

\subsection{\label{subsec:Methods:Model} Reflectance Model}
For simplicity we use a classical model of stack reflectivity based on a Lorentzian susceptibility for the $\mathrm{MoSe_2}$ exciton, taking into account radiative broadening in vacuum $\gamma_{r,0}$ and non-radiative broadening $\gamma_{nr}$ \cite{WootenOpticalPropertiesOfSolids, LargeExcitonicReflectivity}:
\begin{equation}
\chi_{\text{exc}} = - \frac{c}{\omega_0 d} \frac{\gamma_{r,0}}{\omega - \omega_0 + \frac{i \gamma_{nr}}{2}}
\label{eq:exciton_chi}
\end{equation}
where $\omega_0$ is the exciton center frequency, $\omega$ is the optical frequency, $c$ is the speed of light, and $d$ is the $\mathrm{MoSe_2}$ thickness.
The index of refraction of the $\mathrm{MoSe_2}$ is then:
\begin{equation}
n_\text{exc} = \sqrt{n_0^2 + \chi_{\text{exc}}}
\label{eq:index}
\end{equation}
where $n_0$ is the background index in the $\mathrm{MoSe_2}$.
Reflectance from the full stack $R_{\omega_0}(\omega)$ including the mirror is calculated using a transfer-matrix-method to obtain Fresnel coefficients \cite{TranferMatrixMethod}.
Inhomogeneous broadening effects are included with a characteristic width of $\gamma_{ib}$.
To obtain the reflectance $R(\omega)$ including inhomogeneous broadening, $R_{\omega_0}(\omega)$ is calculated for a range of exciton center frequencies and combined by weighting with a Gaussian of width $\gamma_{ib}$:
\begin{equation}
R(\omega) = \frac{1}{\sqrt{2 \pi \gamma_{ib}}} \int{R_{\omega_0^\prime}(\omega) e^{-(\omega_0-\omega^\prime_0)^2/2\gamma_{ib}^2} \mathop{d\omega^\prime_0}}
\label{eq:gamma_ib}
\end{equation}
This assumes that the inhomogeneously broadened excitons emit incoherently, so that interference effects average out.

For fitting purposes, data from several line cuts at characteristic mirror positions were selected.
The model was fit to all line cuts simultaneously, resulting in the following values for the free parameters: $\omega_0=1647.72$ meV, $\gamma_{r, 0} = 1.09$ meV, $\gamma_{nr}=0.40$ meV and $\gamma_{ib}=0.26$ meV.
The mirror distance $z$ was also extracted from the spectra by fitting interference fringes at wavelengths far from $\omega_0$.
See the supplemental information for more details of both fitting procedures, and the values for the static parameters used in the reflectance model.

To extract the model parameters in Fig. \ref{fig:Parameters:ExtractedLinewidth} we first extract the total FWHM linewidth of the $X_0$ feature from the reflectance model as a function of mirror position $z$.
Assuming a Voight line shape \cite{VoightBriefReview, EmpiricalApproximationVoight} we can extract the intrinsic Lorentzian linewidth ($\gamma_r + \gamma_{nr}$) because the intrinsic Gaussian linewidth ($\gamma_{ib}$) is known directly from the model.
This also yields the effective contribution to the linewidth of inhomogeneous broadening ($\gamma_{ib, \mathrm{eff}}$) as the difference between the total linewidth and the intrinsic Lorentzian linewidth.
We can then trivially extract $\gamma_r$ from the intrinsic Lorentzian linewidth because $\gamma_{nr}$ is known.

\subsection{\label{subsec:Methods:Linewidths} Simplified Linewidth Model}
Simplified models for the linewidth modulation of 0D and 2D dipoles parallel to a perfect planar mirror were derived.
The normalized total linewidth as a function of distance from the mirror is \cite{Drexhage1970}:
\begin{equation}
\frac{\gamma_\mathrm{tot}(x)}{\gamma_{\mathrm{tot}, 0}} = 1 + \eta_0 \left[\frac{3\sin x}{2x^3}- \frac{3\cos x}{2x^2} -\frac{3\sin x}{2x} \right]
\label{eq:drexhage_linewidth}
\end{equation}
where $\gamma_\mathrm{tot}(x)$ is the total linewidth at normalized mirror position $x = \frac{4 \pi z}{\lambda_0}$, $\gamma_{\mathrm{tot}, 0}$ is the total linewidth in vacuum and $\lambda_0$ is the wavelength in vacuum.
Following a similar procedure an equation for a 2D material was derived:
\begin{equation}
\frac{\gamma_\mathrm{tot}(x)}{\gamma_{\mathrm{tot}, 0}} = 1 - \eta_0 \cos x
\label{eq:2D_linewidth}
\end{equation}
See the supplemental information for details of the derivation.


%
%

\bibliography{CoherentControl}

\providecommand{\noopsort}[1]{}\providecommand{\singleletter}[1]{#1}%
\begin{thebibliography}{49}%
\makeatletter
\providecommand \@ifxundefined [1]{%
 \@ifx{#1\undefined}
}%
\providecommand \@ifnum [1]{%
 \ifnum #1\expandafter \@firstoftwo
 \else \expandafter \@secondoftwo
 \fi
}%
\providecommand \@ifx [1]{%
 \ifx #1\expandafter \@firstoftwo
 \else \expandafter \@secondoftwo
 \fi
}%
\providecommand \natexlab [1]{#1}%
\providecommand \enquote  [1]{``#1''}%
\providecommand \bibnamefont  [1]{#1}%
\providecommand \bibfnamefont [1]{#1}%
\providecommand \citenamefont [1]{#1}%
\providecommand \href@noop [0]{\@secondoftwo}%
\providecommand \href [0]{\begingroup \@sanitize@url \@href}%
\providecommand \@href[1]{\@@startlink{#1}\@@href}%
\providecommand \@@href[1]{\endgroup#1\@@endlink}%
\providecommand \@sanitize@url [0]{\catcode `\\12\catcode `\$12\catcode
  `\&12\catcode `\#12\catcode `\^12\catcode `\_12\catcode `\%12\relax}%
\providecommand \@@startlink[1]{}%
\providecommand \@@endlink[0]{}%
\providecommand \url  [0]{\begingroup\@sanitize@url \@url }%
\providecommand \@url [1]{\endgroup\@href {#1}{\urlprefix }}%
\providecommand \urlprefix  [0]{URL }%
\providecommand \Eprint [0]{\href }%
\providecommand \doibase [0]{http://dx.doi.org/}%
\providecommand \selectlanguage [0]{\@gobble}%
\providecommand \bibinfo  [0]{\@secondoftwo}%
\providecommand \bibfield  [0]{\@secondoftwo}%
\providecommand \translation [1]{[#1]}%
\providecommand \BibitemOpen [0]{}%
\providecommand \bibitemStop [0]{}%
\providecommand \bibitemNoStop [0]{.\EOS\space}%
\providecommand \EOS [0]{\spacefactor3000\relax}%
\providecommand \BibitemShut  [1]{\csname bibitem#1\endcsname}%
\let\auto@bib@innerbib\@empty
\bibitem [{\citenamefont {Purcell}\ \emph {et~al.}(1946)\citenamefont
  {Purcell}, \citenamefont {Torrey},\ and\ \citenamefont
  {Pound}}]{Purcell_Effect}%
  \BibitemOpen
  \bibfield  {author} {\bibinfo {author} {\bibfnamefont {E.~M.}\ \bibnamefont
  {Purcell}}, \bibinfo {author} {\bibfnamefont {H.~C.}\ \bibnamefont {Torrey}},
  \ and\ \bibinfo {author} {\bibfnamefont {R.~V.}\ \bibnamefont {Pound}},\
  }\bibfield  {title} {\enquote {\bibinfo {title} {{Resonance Absorption by
  Nuclear Magnetic Moments in a Solid}},}\ }\href {\doibase
  10.1103/PhysRev.69.37} {\bibfield  {journal} {\bibinfo  {journal} {Phys.
  Rev.}\ }\textbf {\bibinfo {volume} {69}},\ \bibinfo {pages} {37--38}
  (\bibinfo {year} {1946})}\BibitemShut {NoStop}%
\bibitem [{\citenamefont {Heinzen}\ \emph {et~al.}(1987)\citenamefont
  {Heinzen}, \citenamefont {Childs}, \citenamefont {Thomas},\ and\
  \citenamefont
  {Feld}}]{EnhancedAndInhibitedVisibleSpontaneousEMissionByAtoms}%
  \BibitemOpen
  \bibfield  {author} {\bibinfo {author} {\bibfnamefont {D.~J.}\ \bibnamefont
  {Heinzen}}, \bibinfo {author} {\bibfnamefont {J.~J.}\ \bibnamefont {Childs}},
  \bibinfo {author} {\bibfnamefont {J.~E.}\ \bibnamefont {Thomas}}, \ and\
  \bibinfo {author} {\bibfnamefont {M.~S.}\ \bibnamefont {Feld}},\ }\bibfield
  {title} {\enquote {\bibinfo {title} {{Enhanced and inhibited visible
  spontaneous emission by atoms in a confocal resonator}},}\ }\href {\doibase
  10.1103/PhysRevLett.58.1320} {\bibfield  {journal} {\bibinfo  {journal}
  {Phys. Rev. Lett.}\ }\textbf {\bibinfo {volume} {58}},\ \bibinfo {pages}
  {1320--1323} (\bibinfo {year} {1987})}\BibitemShut {NoStop}%
\bibitem [{\citenamefont {Vredenberg}\ \emph {et~al.}(1993)\citenamefont
  {Vredenberg}, \citenamefont {Hunt}, \citenamefont {Schubert}, \citenamefont
  {Jacobson}, \citenamefont {Poate},\ and\ \citenamefont
  {Zydzik}}]{ControlledAtomicSpontaneousEmission}%
  \BibitemOpen
  \bibfield  {author} {\bibinfo {author} {\bibfnamefont {A.~M.}\ \bibnamefont
  {Vredenberg}}, \bibinfo {author} {\bibfnamefont {N.~E.~J.}\ \bibnamefont
  {Hunt}}, \bibinfo {author} {\bibfnamefont {E.~F.}\ \bibnamefont {Schubert}},
  \bibinfo {author} {\bibfnamefont {D.~C.}\ \bibnamefont {Jacobson}}, \bibinfo
  {author} {\bibfnamefont {J.~M.}\ \bibnamefont {Poate}}, \ and\ \bibinfo
  {author} {\bibfnamefont {G.~J.}\ \bibnamefont {Zydzik}},\ }\bibfield  {title}
  {\enquote {\bibinfo {title} {{Controlled atomic spontaneous emission from
  ${\mathrm{Er}}^{3+}$ in a transparent Si/${\mathrm{SiO}}_{2}$
  microcavity}},}\ }\href {\doibase 10.1103/PhysRevLett.71.517} {\bibfield
  {journal} {\bibinfo  {journal} {Phys. Rev. Lett.}\ }\textbf {\bibinfo
  {volume} {71}},\ \bibinfo {pages} {517--520} (\bibinfo {year}
  {1993})}\BibitemShut {NoStop}%
\bibitem [{\citenamefont {Drexhage}\ \emph {et~al.}(1968)\citenamefont
  {Drexhage}, \citenamefont {Kuhn},\ and\ \citenamefont
  {Sch\"{a}fer}}]{Drexhage1968}%
  \BibitemOpen
  \bibfield  {author} {\bibinfo {author} {\bibfnamefont {K.~H.}\ \bibnamefont
  {Drexhage}}, \bibinfo {author} {\bibfnamefont {H.}~\bibnamefont {Kuhn}}, \
  and\ \bibinfo {author} {\bibfnamefont {F.~P.}\ \bibnamefont {Sch\"{a}fer}},\
  }\bibfield  {title} {\enquote {\bibinfo {title} {{Variation of the
  Fluorescence Decay Time of a Molecule in Front of a Mirror}},}\ }\href
  {\doibase 10.1002/bbpc.19680720261} {\bibfield  {journal} {\bibinfo
  {journal} {Berichte der Bunsengesellschaft für physikalische Chemie}\
  }\textbf {\bibinfo {volume} {72}},\ \bibinfo {pages} {329--329} (\bibinfo
  {year} {1968})}\BibitemShut {NoStop}%
\bibitem [{\citenamefont {Drexhage}(1970)}]{Drexhage1970}%
  \BibitemOpen
  \bibfield  {author} {\bibinfo {author} {\bibfnamefont {K.H.}\ \bibnamefont
  {Drexhage}},\ }\bibfield  {title} {\enquote {\bibinfo {title} {Influence of a
  dielectric interface on fluorescence decay time},}\ }\href {\doibase
  https://doi.org/10.1016/0022-2313(70)90082-7} {\bibfield  {journal} {\bibinfo
   {journal} {Journal of Luminescence}\ }\textbf {\bibinfo {volume} {1-2}},\
  \bibinfo {pages} {693 -- 701} (\bibinfo {year} {1970})}\BibitemShut {NoStop}%
\bibitem [{\citenamefont {Eschner}\ \emph {et~al.}(2001)\citenamefont
  {Eschner}, \citenamefont {Raab}, \citenamefont {Schmidt-Kaler},\ and\
  \citenamefont {Blatt}}]{Eschner2001_Atoms_Mirror}%
  \BibitemOpen
  \bibfield  {author} {\bibinfo {author} {\bibfnamefont {J.}~\bibnamefont
  {Eschner}}, \bibinfo {author} {\bibfnamefont {Ch}~\bibnamefont {Raab}},
  \bibinfo {author} {\bibfnamefont {F.}~\bibnamefont {Schmidt-Kaler}}, \ and\
  \bibinfo {author} {\bibfnamefont {R.}~\bibnamefont {Blatt}},\ }\bibfield
  {title} {\enquote {\bibinfo {title} {{Light interference from single atoms
  and their mirror images}},}\ }\href {https://doi.org/10.1038/35097017}
  {\bibfield  {journal} {\bibinfo  {journal} {Nature}\ }\textbf {\bibinfo
  {volume} {413}},\ \bibinfo {pages} {495 EP --} (\bibinfo {year}
  {2001})}\BibitemShut {NoStop}%
\bibitem [{\citenamefont {Wilson}\ \emph {et~al.}(2003)\citenamefont {Wilson},
  \citenamefont {Bushev}, \citenamefont {Eschner}, \citenamefont
  {Schmidt-Kaler}, \citenamefont {Becher}, \citenamefont {Blatt},\ and\
  \citenamefont {Dorner}}]{TrappedIonVacuumFieldLevelShifts}%
  \BibitemOpen
  \bibfield  {author} {\bibinfo {author} {\bibfnamefont {M.~A.}\ \bibnamefont
  {Wilson}}, \bibinfo {author} {\bibfnamefont {P.}~\bibnamefont {Bushev}},
  \bibinfo {author} {\bibfnamefont {J.}~\bibnamefont {Eschner}}, \bibinfo
  {author} {\bibfnamefont {F.}~\bibnamefont {Schmidt-Kaler}}, \bibinfo {author}
  {\bibfnamefont {C.}~\bibnamefont {Becher}}, \bibinfo {author} {\bibfnamefont
  {R.}~\bibnamefont {Blatt}}, \ and\ \bibinfo {author} {\bibfnamefont
  {U.}~\bibnamefont {Dorner}},\ }\bibfield  {title} {\enquote {\bibinfo {title}
  {{Vacuum-Field Level Shifts in a Single Trapped Ion Mediated by a Single
  Distant Mirror}},}\ }\href {\doibase 10.1103/PhysRevLett.91.213602}
  {\bibfield  {journal} {\bibinfo  {journal} {Phys. Rev. Lett.}\ }\textbf
  {\bibinfo {volume} {91}},\ \bibinfo {pages} {213602} (\bibinfo {year}
  {2003})}\BibitemShut {NoStop}%
\bibitem [{\citenamefont {Stobbe}\ \emph {et~al.}(2009)\citenamefont {Stobbe},
  \citenamefont {Johansen}, \citenamefont {Kristensen}, \citenamefont {Hvam},\
  and\ \citenamefont {Lodahl}}]{Stobbe2009_QD_Mirror}%
  \BibitemOpen
  \bibfield  {author} {\bibinfo {author} {\bibfnamefont {S.}~\bibnamefont
  {Stobbe}}, \bibinfo {author} {\bibfnamefont {J.}~\bibnamefont {Johansen}},
  \bibinfo {author} {\bibfnamefont {P.~T.}\ \bibnamefont {Kristensen}},
  \bibinfo {author} {\bibfnamefont {J.~M.}\ \bibnamefont {Hvam}}, \ and\
  \bibinfo {author} {\bibfnamefont {P.}~\bibnamefont {Lodahl}},\ }\bibfield
  {title} {\enquote {\bibinfo {title} {{Frequency dependence of the radiative
  decay rate of excitons in self-assembled quantum dots: Experiment and
  theory}},}\ }\href {\doibase 10.1103/PhysRevB.80.155307} {\bibfield
  {journal} {\bibinfo  {journal} {Phys. Rev. B}\ }\textbf {\bibinfo {volume}
  {80}},\ \bibinfo {pages} {155307} (\bibinfo {year} {2009})}\BibitemShut
  {NoStop}%
\bibitem [{\citenamefont {Cadiz}\ \emph {et~al.}(2017)\citenamefont {Cadiz},
  \citenamefont {Courtade}, \citenamefont {Robert}, \citenamefont {Wang},
  \citenamefont {Shen}, \citenamefont {Cai}, \citenamefont {Taniguchi},
  \citenamefont {Watanabe}, \citenamefont {Carrere}, \citenamefont {Lagarde},
  \citenamefont {Manca}, \citenamefont {Amand}, \citenamefont {Renucci},
  \citenamefont {Tongay}, \citenamefont {Marie},\ and\ \citenamefont
  {Urbaszek}}]{ExcitonicLinewidthApproachingHomogenousLimit}%
  \BibitemOpen
  \bibfield  {author} {\bibinfo {author} {\bibfnamefont {F.}~\bibnamefont
  {Cadiz}}, \bibinfo {author} {\bibfnamefont {E.}~\bibnamefont {Courtade}},
  \bibinfo {author} {\bibfnamefont {C.}~\bibnamefont {Robert}}, \bibinfo
  {author} {\bibfnamefont {G.}~\bibnamefont {Wang}}, \bibinfo {author}
  {\bibfnamefont {Y.}~\bibnamefont {Shen}}, \bibinfo {author} {\bibfnamefont
  {H.}~\bibnamefont {Cai}}, \bibinfo {author} {\bibfnamefont {T.}~\bibnamefont
  {Taniguchi}}, \bibinfo {author} {\bibfnamefont {K.}~\bibnamefont {Watanabe}},
  \bibinfo {author} {\bibfnamefont {H.}~\bibnamefont {Carrere}}, \bibinfo
  {author} {\bibfnamefont {D.}~\bibnamefont {Lagarde}}, \bibinfo {author}
  {\bibfnamefont {M.}~\bibnamefont {Manca}}, \bibinfo {author} {\bibfnamefont
  {T.}~\bibnamefont {Amand}}, \bibinfo {author} {\bibfnamefont
  {P.}~\bibnamefont {Renucci}}, \bibinfo {author} {\bibfnamefont
  {S.}~\bibnamefont {Tongay}}, \bibinfo {author} {\bibfnamefont
  {X.}~\bibnamefont {Marie}}, \ and\ \bibinfo {author} {\bibfnamefont
  {B.}~\bibnamefont {Urbaszek}},\ }\bibfield  {title} {\enquote {\bibinfo
  {title} {{Excitonic Linewidth Approaching the Homogeneous Limit in
  ${\mathrm{MoS}}_{2}$-Based van der Waals Heterostructures}},}\ }\href
  {\doibase 10.1103/PhysRevX.7.021026} {\bibfield  {journal} {\bibinfo
  {journal} {Phys. Rev. X}\ }\textbf {\bibinfo {volume} {7}},\ \bibinfo {pages}
  {021026} (\bibinfo {year} {2017})}\BibitemShut {NoStop}%
\bibitem [{\citenamefont {Scuri}\ \emph {et~al.}(2018)\citenamefont {Scuri},
  \citenamefont {Zhou}, \citenamefont {High}, \citenamefont {Wild},
  \citenamefont {Shu}, \citenamefont {De~Greve}, \citenamefont {Jauregui},
  \citenamefont {Taniguchi}, \citenamefont {Watanabe}, \citenamefont {Kim},
  \citenamefont {Lukin},\ and\ \citenamefont
  {Park}}]{LargeExcitonicReflectivity}%
  \BibitemOpen
  \bibfield  {author} {\bibinfo {author} {\bibfnamefont {Giovanni}\
  \bibnamefont {Scuri}}, \bibinfo {author} {\bibfnamefont {You}\ \bibnamefont
  {Zhou}}, \bibinfo {author} {\bibfnamefont {Alexander~A.}\ \bibnamefont
  {High}}, \bibinfo {author} {\bibfnamefont {Dominik~S.}\ \bibnamefont {Wild}},
  \bibinfo {author} {\bibfnamefont {Chi}\ \bibnamefont {Shu}}, \bibinfo
  {author} {\bibfnamefont {Kristiaan}\ \bibnamefont {De~Greve}}, \bibinfo
  {author} {\bibfnamefont {Luis~A.}\ \bibnamefont {Jauregui}}, \bibinfo
  {author} {\bibfnamefont {Takashi}\ \bibnamefont {Taniguchi}}, \bibinfo
  {author} {\bibfnamefont {Kenji}\ \bibnamefont {Watanabe}}, \bibinfo {author}
  {\bibfnamefont {Philip}\ \bibnamefont {Kim}}, \bibinfo {author}
  {\bibfnamefont {Mikhail~D.}\ \bibnamefont {Lukin}}, \ and\ \bibinfo {author}
  {\bibfnamefont {Hongkun}\ \bibnamefont {Park}},\ }\bibfield  {title}
  {\enquote {\bibinfo {title} {{Large Excitonic Reflectivity of Monolayer
  ${\mathrm{MoSe}}_{2}$ Encapsulated in Hexagonal Boron Nitride}},}\ }\href
  {\doibase 10.1103/PhysRevLett.120.037402} {\bibfield  {journal} {\bibinfo
  {journal} {Phys. Rev. Lett.}\ }\textbf {\bibinfo {volume} {120}},\ \bibinfo
  {pages} {037402} (\bibinfo {year} {2018})}\BibitemShut {NoStop}%
\bibitem [{\citenamefont {Back}\ \emph {et~al.}(2018)\citenamefont {Back},
  \citenamefont {Zeytinoglu}, \citenamefont {Ijaz}, \citenamefont {Kroner},\
  and\ \citenamefont {Imamo\ifmmode~\breve{g}\else
  \u{g}\fi{}lu}}]{RealizationOfAnElectricallyTunableMirror}%
  \BibitemOpen
  \bibfield  {author} {\bibinfo {author} {\bibfnamefont {Patrick}\ \bibnamefont
  {Back}}, \bibinfo {author} {\bibfnamefont {Sina}\ \bibnamefont {Zeytinoglu}},
  \bibinfo {author} {\bibfnamefont {Aroosa}\ \bibnamefont {Ijaz}}, \bibinfo
  {author} {\bibfnamefont {Martin}\ \bibnamefont {Kroner}}, \ and\ \bibinfo
  {author} {\bibfnamefont {Atac}\ \bibnamefont {Imamo\ifmmode~\breve{g}\else
  \u{g}\fi{}lu}},\ }\bibfield  {title} {\enquote {\bibinfo {title}
  {{Realization of an Electrically Tunable Narrow-Bandwidth Atomically Thin
  Mirror Using Monolayer ${\mathrm{MoSe}}_{2}$}},}\ }\href {\doibase
  10.1103/PhysRevLett.120.037401} {\bibfield  {journal} {\bibinfo  {journal}
  {Phys. Rev. Lett.}\ }\textbf {\bibinfo {volume} {120}},\ \bibinfo {pages}
  {037401} (\bibinfo {year} {2018})}\BibitemShut {NoStop}%
\bibitem [{\citenamefont {Meschede}\ \emph {et~al.}(1990)\citenamefont
  {Meschede}, \citenamefont {Jhe},\ and\ \citenamefont
  {Hinds}}]{RadiativePropertiesofAtomsNearAConductingPlane}%
  \BibitemOpen
  \bibfield  {author} {\bibinfo {author} {\bibfnamefont {D.}~\bibnamefont
  {Meschede}}, \bibinfo {author} {\bibfnamefont {W.}~\bibnamefont {Jhe}}, \
  and\ \bibinfo {author} {\bibfnamefont {E.~A.}\ \bibnamefont {Hinds}},\
  }\bibfield  {title} {\enquote {\bibinfo {title} {{Radiative properties of
  atoms near a conducting plane: An old problem in a new light}},}\ }\href
  {\doibase 10.1103/PhysRevA.41.1587} {\bibfield  {journal} {\bibinfo
  {journal} {Phys. Rev. A}\ }\textbf {\bibinfo {volume} {41}},\ \bibinfo
  {pages} {1587--1596} (\bibinfo {year} {1990})}\BibitemShut {NoStop}%
\bibitem [{\citenamefont {Byrnes}\ \emph {et~al.}(2014)\citenamefont {Byrnes},
  \citenamefont {Kim},\ and\ \citenamefont
  {Yamamoto}}]{ExcitonPolaritonCondensateReview}%
  \BibitemOpen
  \bibfield  {author} {\bibinfo {author} {\bibfnamefont {Tim}\ \bibnamefont
  {Byrnes}}, \bibinfo {author} {\bibfnamefont {Na~Young}\ \bibnamefont {Kim}},
  \ and\ \bibinfo {author} {\bibfnamefont {Yoshihisa}\ \bibnamefont
  {Yamamoto}},\ }\bibfield  {title} {\enquote {\bibinfo {title}
  {{Exciton-polariton condensates}},}\ }\href
  {https://doi.org/10.1038/nphys3143} {\bibfield  {journal} {\bibinfo
  {journal} {Nature Physics}\ }\textbf {\bibinfo {volume} {10}},\ \bibinfo
  {pages} {803 EP --} (\bibinfo {year} {2014})},\ \bibinfo {note} {review
  Article}\BibitemShut {NoStop}%
\bibitem [{\citenamefont {Zeytino\ifmmode~\check{g}\else \v{g}\fi{}lu}\ \emph
  {et~al.}(2017)\citenamefont {Zeytino\ifmmode~\check{g}\else \v{g}\fi{}lu},
  \citenamefont {Roth}, \citenamefont {Huber},\ and\ \citenamefont {\ifmmode
  \dot{I}\else \.{I}\fi{}mamo\ifmmode~\breve{g}\else
  \u{g}\fi{}lu}}]{Imamoglu_PRA_NonlinearMirror}%
  \BibitemOpen
  \bibfield  {author} {\bibinfo {author} {\bibfnamefont {Sina}\ \bibnamefont
  {Zeytino\ifmmode~\check{g}\else \v{g}\fi{}lu}}, \bibinfo {author}
  {\bibfnamefont {Charlaine}\ \bibnamefont {Roth}}, \bibinfo {author}
  {\bibfnamefont {Sebastian}\ \bibnamefont {Huber}}, \ and\ \bibinfo {author}
  {\bibfnamefont {Atac}\ \bibnamefont {\ifmmode \dot{I}\else
  \.{I}\fi{}mamo\ifmmode~\breve{g}\else \u{g}\fi{}lu}},\ }\bibfield  {title}
  {\enquote {\bibinfo {title} {{Atomically thin semiconductors as nonlinear
  mirrors}},}\ }\href {\doibase 10.1103/PhysRevA.96.031801} {\bibfield
  {journal} {\bibinfo  {journal} {Phys. Rev. A}\ }\textbf {\bibinfo {volume}
  {96}},\ \bibinfo {pages} {031801} (\bibinfo {year} {2017})}\BibitemShut
  {NoStop}%
\bibitem [{\citenamefont {Wild}\ \emph {et~al.}(2018)\citenamefont {Wild},
  \citenamefont {Shahmoon}, \citenamefont {Yelin},\ and\ \citenamefont
  {Lukin}}]{Wild_PRL_QuantumNonlinearOptics}%
  \BibitemOpen
  \bibfield  {author} {\bibinfo {author} {\bibfnamefont {Dominik~S.}\
  \bibnamefont {Wild}}, \bibinfo {author} {\bibfnamefont {Ephraim}\
  \bibnamefont {Shahmoon}}, \bibinfo {author} {\bibfnamefont {Susanne~F.}\
  \bibnamefont {Yelin}}, \ and\ \bibinfo {author} {\bibfnamefont {Mikhail~D.}\
  \bibnamefont {Lukin}},\ }\bibfield  {title} {\enquote {\bibinfo {title}
  {{Quantum Nonlinear Optics in Atomically Thin Materials}},}\ }\href {\doibase
  10.1103/PhysRevLett.121.123606} {\bibfield  {journal} {\bibinfo  {journal}
  {Phys. Rev. Lett.}\ }\textbf {\bibinfo {volume} {121}},\ \bibinfo {pages}
  {123606} (\bibinfo {year} {2018})}\BibitemShut {NoStop}%
\bibitem [{\citenamefont {Mak}\ \emph {et~al.}(2010)\citenamefont {Mak},
  \citenamefont {Lee}, \citenamefont {Hone}, \citenamefont {Shan},\ and\
  \citenamefont {Heinz}}]{AtomicallyThinMoS2}%
  \BibitemOpen
  \bibfield  {author} {\bibinfo {author} {\bibfnamefont {Kin~Fai}\ \bibnamefont
  {Mak}}, \bibinfo {author} {\bibfnamefont {Changgu}\ \bibnamefont {Lee}},
  \bibinfo {author} {\bibfnamefont {James}\ \bibnamefont {Hone}}, \bibinfo
  {author} {\bibfnamefont {Jie}\ \bibnamefont {Shan}}, \ and\ \bibinfo {author}
  {\bibfnamefont {Tony~F.}\ \bibnamefont {Heinz}},\ }\bibfield  {title}
  {\enquote {\bibinfo {title} {{Atomically Thin ${\mathrm{MoS}}_{2}$: A New
  Direct-Gap Semiconductor}},}\ }\href {\doibase
  10.1103/PhysRevLett.105.136805} {\bibfield  {journal} {\bibinfo  {journal}
  {Phys. Rev. Lett.}\ }\textbf {\bibinfo {volume} {105}},\ \bibinfo {pages}
  {136805} (\bibinfo {year} {2010})}\BibitemShut {NoStop}%
\bibitem [{\citenamefont {Splendiani}\ \emph {et~al.}(2010)\citenamefont
  {Splendiani}, \citenamefont {Sun}, \citenamefont {Zhang}, \citenamefont {Li},
  \citenamefont {Kim}, \citenamefont {Chim}, \citenamefont {Galli},\ and\
  \citenamefont {Wang}}]{EmergingPhotoluminescence}%
  \BibitemOpen
  \bibfield  {author} {\bibinfo {author} {\bibfnamefont {Andrea}\ \bibnamefont
  {Splendiani}}, \bibinfo {author} {\bibfnamefont {Liang}\ \bibnamefont {Sun}},
  \bibinfo {author} {\bibfnamefont {Yuanbo}\ \bibnamefont {Zhang}}, \bibinfo
  {author} {\bibfnamefont {Tianshu}\ \bibnamefont {Li}}, \bibinfo {author}
  {\bibfnamefont {Jonghwan}\ \bibnamefont {Kim}}, \bibinfo {author}
  {\bibfnamefont {Chi-Yung}\ \bibnamefont {Chim}}, \bibinfo {author}
  {\bibfnamefont {Giulia}\ \bibnamefont {Galli}}, \ and\ \bibinfo {author}
  {\bibfnamefont {Feng}\ \bibnamefont {Wang}},\ }\bibfield  {title} {\enquote
  {\bibinfo {title} {{Emerging Photoluminescence in Monolayer
  $\mathrm{MoS_2}$}},}\ }\href {\doibase 10.1021/nl903868w} {\bibfield
  {journal} {\bibinfo  {journal} {Nano Letters}\ }\textbf {\bibinfo {volume}
  {10}},\ \bibinfo {pages} {1271--1275} (\bibinfo {year} {2010})}\BibitemShut
  {NoStop}%
\bibitem [{\citenamefont {Zhang}\ \emph {et~al.}(2013)\citenamefont {Zhang},
  \citenamefont {Chang}, \citenamefont {Zhou}, \citenamefont {Cui},
  \citenamefont {Yan}, \citenamefont {Liu}, \citenamefont {Schmitt},
  \citenamefont {Lee}, \citenamefont {Moore}, \citenamefont {Chen},
  \citenamefont {Lin}, \citenamefont {Jeng}, \citenamefont {Mo}, \citenamefont
  {Hussain}, \citenamefont {Bansil},\ and\ \citenamefont
  {Shen}}]{MoSe2DirectBandgap}%
  \BibitemOpen
  \bibfield  {author} {\bibinfo {author} {\bibfnamefont {Yi}~\bibnamefont
  {Zhang}}, \bibinfo {author} {\bibfnamefont {Tay-Rong}\ \bibnamefont {Chang}},
  \bibinfo {author} {\bibfnamefont {Bo}~\bibnamefont {Zhou}}, \bibinfo {author}
  {\bibfnamefont {Yong-Tao}\ \bibnamefont {Cui}}, \bibinfo {author}
  {\bibfnamefont {Hao}\ \bibnamefont {Yan}}, \bibinfo {author} {\bibfnamefont
  {Zhongkai}\ \bibnamefont {Liu}}, \bibinfo {author} {\bibfnamefont {Felix}\
  \bibnamefont {Schmitt}}, \bibinfo {author} {\bibfnamefont {James}\
  \bibnamefont {Lee}}, \bibinfo {author} {\bibfnamefont {Rob}\ \bibnamefont
  {Moore}}, \bibinfo {author} {\bibfnamefont {Yulin}\ \bibnamefont {Chen}},
  \bibinfo {author} {\bibfnamefont {Hsin}\ \bibnamefont {Lin}}, \bibinfo
  {author} {\bibfnamefont {Horng-Tay}\ \bibnamefont {Jeng}}, \bibinfo {author}
  {\bibfnamefont {Sung-Kwan}\ \bibnamefont {Mo}}, \bibinfo {author}
  {\bibfnamefont {Zahid}\ \bibnamefont {Hussain}}, \bibinfo {author}
  {\bibfnamefont {Arun}\ \bibnamefont {Bansil}}, \ and\ \bibinfo {author}
  {\bibfnamefont {Zhi-Xun}\ \bibnamefont {Shen}},\ }\bibfield  {title}
  {\enquote {\bibinfo {title} {{Direct observation of the transition from
  indirect to direct bandgap in atomically thin epitaxial $\mathrm{MoSe2}$}},}\
  }\href {https://doi.org/10.1038/nnano.2013.277} {\bibfield  {journal}
  {\bibinfo  {journal} {Nature Nanotechnology}\ }\textbf {\bibinfo {volume}
  {9}},\ \bibinfo {pages} {111 EP --} (\bibinfo {year} {2013})}\BibitemShut
  {NoStop}%
\bibitem [{\citenamefont {Molina-S\'anchez}\ \emph {et~al.}(2013)\citenamefont
  {Molina-S\'anchez}, \citenamefont {Sangalli}, \citenamefont {Hummer},
  \citenamefont {Marini},\ and\ \citenamefont {Wirtz}}]{PRB2013_MoS2AbInitio}%
  \BibitemOpen
  \bibfield  {author} {\bibinfo {author} {\bibfnamefont {Alejandro}\
  \bibnamefont {Molina-S\'anchez}}, \bibinfo {author} {\bibfnamefont {Davide}\
  \bibnamefont {Sangalli}}, \bibinfo {author} {\bibfnamefont {Kerstin}\
  \bibnamefont {Hummer}}, \bibinfo {author} {\bibfnamefont {Andrea}\
  \bibnamefont {Marini}}, \ and\ \bibinfo {author} {\bibfnamefont {Ludger}\
  \bibnamefont {Wirtz}},\ }\bibfield  {title} {\enquote {\bibinfo {title}
  {{Effect of spin-orbit interaction on the optical spectra of single-layer,
  double-layer, and bulk MoS${}_{2}$}},}\ }\href {\doibase
  10.1103/PhysRevB.88.045412} {\bibfield  {journal} {\bibinfo  {journal} {Phys.
  Rev. B}\ }\textbf {\bibinfo {volume} {88}},\ \bibinfo {pages} {045412}
  (\bibinfo {year} {2013})}\BibitemShut {NoStop}%
\bibitem [{\citenamefont {Molina-S\'anchez}\ \emph {et~al.}(2016)\citenamefont
  {Molina-S\'anchez}, \citenamefont {Palummo}, \citenamefont {Marini},\ and\
  \citenamefont {Wirtz}}]{PRB2016_MoS2_abInitio}%
  \BibitemOpen
  \bibfield  {author} {\bibinfo {author} {\bibfnamefont {Alejandro}\
  \bibnamefont {Molina-S\'anchez}}, \bibinfo {author} {\bibfnamefont
  {Maurizia}\ \bibnamefont {Palummo}}, \bibinfo {author} {\bibfnamefont
  {Andrea}\ \bibnamefont {Marini}}, \ and\ \bibinfo {author} {\bibfnamefont
  {Ludger}\ \bibnamefont {Wirtz}},\ }\bibfield  {title} {\enquote {\bibinfo
  {title} {{Temperature-dependent excitonic effects in the optical properties
  of single-layer ${\mathrm{MoS}}_{2}$}},}\ }\href {\doibase
  10.1103/PhysRevB.93.155435} {\bibfield  {journal} {\bibinfo  {journal} {Phys.
  Rev. B}\ }\textbf {\bibinfo {volume} {93}},\ \bibinfo {pages} {155435}
  (\bibinfo {year} {2016})}\BibitemShut {NoStop}%
\bibitem [{\citenamefont {Jones}\ \emph {et~al.}(2013)\citenamefont {Jones},
  \citenamefont {Yu}, \citenamefont {Ghimire}, \citenamefont {Wu},
  \citenamefont {Aivazian}, \citenamefont {Ross}, \citenamefont {Zhao},
  \citenamefont {Yan}, \citenamefont {Mandrus}, \citenamefont {Xiao},
  \citenamefont {Yao},\ and\ \citenamefont {Xu}}]{ExcitonValleyCoherence}%
  \BibitemOpen
  \bibfield  {author} {\bibinfo {author} {\bibfnamefont {Aaron~M.}\
  \bibnamefont {Jones}}, \bibinfo {author} {\bibfnamefont {Hongyi}\
  \bibnamefont {Yu}}, \bibinfo {author} {\bibfnamefont {Nirmal~J.}\
  \bibnamefont {Ghimire}}, \bibinfo {author} {\bibfnamefont {Sanfeng}\
  \bibnamefont {Wu}}, \bibinfo {author} {\bibfnamefont {Grant}\ \bibnamefont
  {Aivazian}}, \bibinfo {author} {\bibfnamefont {Jason~S.}\ \bibnamefont
  {Ross}}, \bibinfo {author} {\bibfnamefont {Bo}~\bibnamefont {Zhao}}, \bibinfo
  {author} {\bibfnamefont {Jiaqiang}\ \bibnamefont {Yan}}, \bibinfo {author}
  {\bibfnamefont {David~G.}\ \bibnamefont {Mandrus}}, \bibinfo {author}
  {\bibfnamefont {Di}~\bibnamefont {Xiao}}, \bibinfo {author} {\bibfnamefont
  {Wang}\ \bibnamefont {Yao}}, \ and\ \bibinfo {author} {\bibfnamefont
  {Xiaodong}\ \bibnamefont {Xu}},\ }\bibfield  {title} {\enquote {\bibinfo
  {title} {{Optical generation of excitonic valley coherence in monolayer
  $\mathrm{WSe_2}$}},}\ }\href {http://dx.doi.org/10.1038/nnano.2013.151}
  {\bibfield  {journal} {\bibinfo  {journal} {Nature Nanotechnology}\ }\textbf
  {\bibinfo {volume} {8}},\ \bibinfo {pages} {634 EP --} (\bibinfo {year}
  {2013})}\BibitemShut {NoStop}%
\bibitem [{\citenamefont {Xiao}\ \emph {et~al.}(2012)\citenamefont {Xiao},
  \citenamefont {Liu}, \citenamefont {Feng}, \citenamefont {Xu},\ and\
  \citenamefont {Yao}}]{CoupledSpinValleyPhysics}%
  \BibitemOpen
  \bibfield  {author} {\bibinfo {author} {\bibfnamefont {Di}~\bibnamefont
  {Xiao}}, \bibinfo {author} {\bibfnamefont {Gui-Bin}\ \bibnamefont {Liu}},
  \bibinfo {author} {\bibfnamefont {Wanxiang}\ \bibnamefont {Feng}}, \bibinfo
  {author} {\bibfnamefont {Xiaodong}\ \bibnamefont {Xu}}, \ and\ \bibinfo
  {author} {\bibfnamefont {Wang}\ \bibnamefont {Yao}},\ }\bibfield  {title}
  {\enquote {\bibinfo {title} {{Coupled Spin and Valley Physics in Monolayers
  of ${\mathrm{MoS}}_{2}$ and Other Group-VI Dichalcogenides}},}\ }\href
  {\doibase 10.1103/PhysRevLett.108.196802} {\bibfield  {journal} {\bibinfo
  {journal} {Phys. Rev. Lett.}\ }\textbf {\bibinfo {volume} {108}},\ \bibinfo
  {pages} {196802} (\bibinfo {year} {2012})}\BibitemShut {NoStop}%
\bibitem [{\citenamefont {Mak}\ \emph {et~al.}(2012{\natexlab{a}})\citenamefont
  {Mak}, \citenamefont {He}, \citenamefont {Shan},\ and\ \citenamefont
  {Heinz}}]{ControlOfValleyPolarization}%
  \BibitemOpen
  \bibfield  {author} {\bibinfo {author} {\bibfnamefont {Kin~Fai}\ \bibnamefont
  {Mak}}, \bibinfo {author} {\bibfnamefont {Keliang}\ \bibnamefont {He}},
  \bibinfo {author} {\bibfnamefont {Jie}\ \bibnamefont {Shan}}, \ and\ \bibinfo
  {author} {\bibfnamefont {Tony~F.}\ \bibnamefont {Heinz}},\ }\bibfield
  {title} {\enquote {\bibinfo {title} {{Control of valley polarization in
  monolayer $\mathrm{MoS_2}$ by optical helicity}},}\ }\href
  {http://dx.doi.org/10.1038/nnano.2012.96} {\bibfield  {journal} {\bibinfo
  {journal} {Nature Nanotechnology}\ }\textbf {\bibinfo {volume} {7}},\
  \bibinfo {pages} {494 EP --} (\bibinfo {year}
  {2012}{\natexlab{a}})}\BibitemShut {NoStop}%
\bibitem [{\citenamefont {Conley}\ \emph {et~al.}(2013)\citenamefont {Conley},
  \citenamefont {Wang}, \citenamefont {Ziegler}, \citenamefont {Haglund},
  \citenamefont {Pantelides},\ and\ \citenamefont
  {Bolotin}}]{BandgapEngineeringOfStrainedMoS2}%
  \BibitemOpen
  \bibfield  {author} {\bibinfo {author} {\bibfnamefont {Hiram~J.}\
  \bibnamefont {Conley}}, \bibinfo {author} {\bibfnamefont {Bin}\ \bibnamefont
  {Wang}}, \bibinfo {author} {\bibfnamefont {Jed~I.}\ \bibnamefont {Ziegler}},
  \bibinfo {author} {\bibfnamefont {Richard~F.}\ \bibnamefont {Haglund}},
  \bibinfo {author} {\bibfnamefont {Sokrates~T.}\ \bibnamefont {Pantelides}}, \
  and\ \bibinfo {author} {\bibfnamefont {Kirill~I.}\ \bibnamefont {Bolotin}},\
  }\bibfield  {title} {\enquote {\bibinfo {title} {{Bandgap Engineering of
  Strained Monolayer and Bilayer $\mathrm{MoS_2}$}},}\ }\href {\doibase
  10.1021/nl4014748} {\bibfield  {journal} {\bibinfo  {journal} {Nano Letters}\
  }\textbf {\bibinfo {volume} {13}},\ \bibinfo {pages} {3626--3630} (\bibinfo
  {year} {2013})}\BibitemShut {NoStop}%
\bibitem [{\citenamefont {You}\ \emph {et~al.}(2015)\citenamefont {You},
  \citenamefont {Zhang}, \citenamefont {Berkelbach}, \citenamefont {Hybertsen},
  \citenamefont {Reichman},\ and\ \citenamefont
  {Heinz}}]{ObservationOfBiexcitonsInMonolayerWSe2}%
  \BibitemOpen
  \bibfield  {author} {\bibinfo {author} {\bibfnamefont {Yumeng}\ \bibnamefont
  {You}}, \bibinfo {author} {\bibfnamefont {Xiao-Xiao}\ \bibnamefont {Zhang}},
  \bibinfo {author} {\bibfnamefont {Timothy~C.}\ \bibnamefont {Berkelbach}},
  \bibinfo {author} {\bibfnamefont {Mark~S.}\ \bibnamefont {Hybertsen}},
  \bibinfo {author} {\bibfnamefont {David~R.}\ \bibnamefont {Reichman}}, \ and\
  \bibinfo {author} {\bibfnamefont {Tony~F.}\ \bibnamefont {Heinz}},\
  }\bibfield  {title} {\enquote {\bibinfo {title} {{Observation of biexcitons
  in monolayer $\mathrm{WSe_2}$}},}\ }\href
  {http://dx.doi.org/10.1038/nphys3324} {\bibfield  {journal} {\bibinfo
  {journal} {Nature Physics}\ }\textbf {\bibinfo {volume} {11}},\ \bibinfo
  {pages} {477 EP --} (\bibinfo {year} {2015})}\BibitemShut {NoStop}%
\bibitem [{\citenamefont {Sidler}\ \emph {et~al.}(2016)\citenamefont {Sidler},
  \citenamefont {Back}, \citenamefont {Cotlet}, \citenamefont {Srivastava},
  \citenamefont {Fink}, \citenamefont {Kroner}, \citenamefont {Demler},\ and\
  \citenamefont {Imamoglu}}]{FermiPolaronPolaritons}%
  \BibitemOpen
  \bibfield  {author} {\bibinfo {author} {\bibfnamefont {Meinrad}\ \bibnamefont
  {Sidler}}, \bibinfo {author} {\bibfnamefont {Patrick}\ \bibnamefont {Back}},
  \bibinfo {author} {\bibfnamefont {Ovidiu}\ \bibnamefont {Cotlet}}, \bibinfo
  {author} {\bibfnamefont {Ajit}\ \bibnamefont {Srivastava}}, \bibinfo {author}
  {\bibfnamefont {Thomas}\ \bibnamefont {Fink}}, \bibinfo {author}
  {\bibfnamefont {Martin}\ \bibnamefont {Kroner}}, \bibinfo {author}
  {\bibfnamefont {Eugene}\ \bibnamefont {Demler}}, \ and\ \bibinfo {author}
  {\bibfnamefont {Atac}\ \bibnamefont {Imamoglu}},\ }\bibfield  {title}
  {\enquote {\bibinfo {title} {{Fermi polaron-polaritons in charge-tunable
  atomically thin semiconductors}},}\ }\href
  {http://dx.doi.org/10.1038/nphys3949} {\bibfield  {journal} {\bibinfo
  {journal} {Nature Physics}\ }\textbf {\bibinfo {volume} {13}},\ \bibinfo
  {pages} {255 EP --} (\bibinfo {year} {2016})}\BibitemShut {NoStop}%
\bibitem [{\citenamefont {Mak}\ \emph {et~al.}(2012{\natexlab{b}})\citenamefont
  {Mak}, \citenamefont {He}, \citenamefont {Lee}, \citenamefont {Lee},
  \citenamefont {Hone}, \citenamefont {Heinz},\ and\ \citenamefont
  {Shan}}]{TightlyBoundTrions}%
  \BibitemOpen
  \bibfield  {author} {\bibinfo {author} {\bibfnamefont {Kin~Fai}\ \bibnamefont
  {Mak}}, \bibinfo {author} {\bibfnamefont {Keliang}\ \bibnamefont {He}},
  \bibinfo {author} {\bibfnamefont {Changgu}\ \bibnamefont {Lee}}, \bibinfo
  {author} {\bibfnamefont {Gwan~Hyoung}\ \bibnamefont {Lee}}, \bibinfo {author}
  {\bibfnamefont {James}\ \bibnamefont {Hone}}, \bibinfo {author}
  {\bibfnamefont {Tony~F.}\ \bibnamefont {Heinz}}, \ and\ \bibinfo {author}
  {\bibfnamefont {Jie}\ \bibnamefont {Shan}},\ }\bibfield  {title} {\enquote
  {\bibinfo {title} {{Tightly bound trions in monolayer $\mathrm{MoS_2}$}},}\
  }\href {http://dx.doi.org/10.1038/nmat3505} {\bibfield  {journal} {\bibinfo
  {journal} {Nature Materials}\ }\textbf {\bibinfo {volume} {12}},\ \bibinfo
  {pages} {207 EP --} (\bibinfo {year} {2012}{\natexlab{b}})}\BibitemShut
  {NoStop}%
\bibitem [{\citenamefont {Wang}\ \emph {et~al.}(2018)\citenamefont {Wang},
  \citenamefont {De~Greve}, \citenamefont {Jauregui}, \citenamefont {Sushko},
  \citenamefont {High}, \citenamefont {Zhou}, \citenamefont {Scuri},
  \citenamefont {Taniguchi}, \citenamefont {Watanabe}, \citenamefont {Lukin},
  \citenamefont {Park},\ and\ \citenamefont
  {Kim}}]{ElectricalControlofChargedCarriersInAtomicallyThin}%
  \BibitemOpen
  \bibfield  {author} {\bibinfo {author} {\bibfnamefont {Ke}~\bibnamefont
  {Wang}}, \bibinfo {author} {\bibfnamefont {Kristiaan}\ \bibnamefont
  {De~Greve}}, \bibinfo {author} {\bibfnamefont {Luis~A.}\ \bibnamefont
  {Jauregui}}, \bibinfo {author} {\bibfnamefont {Andrey}\ \bibnamefont
  {Sushko}}, \bibinfo {author} {\bibfnamefont {Alexander}\ \bibnamefont
  {High}}, \bibinfo {author} {\bibfnamefont {You}\ \bibnamefont {Zhou}},
  \bibinfo {author} {\bibfnamefont {Giovanni}\ \bibnamefont {Scuri}}, \bibinfo
  {author} {\bibfnamefont {Takashi}\ \bibnamefont {Taniguchi}}, \bibinfo
  {author} {\bibfnamefont {Kenji}\ \bibnamefont {Watanabe}}, \bibinfo {author}
  {\bibfnamefont {Mikhail~D.}\ \bibnamefont {Lukin}}, \bibinfo {author}
  {\bibfnamefont {Hongkun}\ \bibnamefont {Park}}, \ and\ \bibinfo {author}
  {\bibfnamefont {Philip}\ \bibnamefont {Kim}},\ }\bibfield  {title} {\enquote
  {\bibinfo {title} {{Electrical control of charged carriers and excitons in
  atomically thin materials}},}\ }\href {\doibase 10.1038/s41565-017-0030-x}
  {\bibfield  {journal} {\bibinfo  {journal} {Nature Nanotechnology}\ }\textbf
  {\bibinfo {volume} {13}},\ \bibinfo {pages} {128--132} (\bibinfo {year}
  {2018})}\BibitemShut {NoStop}%
\bibitem [{\citenamefont {Palacios-Berraquero}\ \emph
  {et~al.}(2017)\citenamefont {Palacios-Berraquero}, \citenamefont {Kara},
  \citenamefont {Montblanch}, \citenamefont {Barbone}, \citenamefont
  {Latawiec}, \citenamefont {Yoon}, \citenamefont {Ott}, \citenamefont
  {Loncar}, \citenamefont {Ferrari},\ and\ \citenamefont
  {Atat{\"u}re}}]{LargeScaleQuantumEmitterArrays}%
  \BibitemOpen
  \bibfield  {author} {\bibinfo {author} {\bibfnamefont {Carmen}\ \bibnamefont
  {Palacios-Berraquero}}, \bibinfo {author} {\bibfnamefont {Dhiren~M.}\
  \bibnamefont {Kara}}, \bibinfo {author} {\bibfnamefont {Alejandro R.~P}\
  \bibnamefont {Montblanch}}, \bibinfo {author} {\bibfnamefont {Matteo}\
  \bibnamefont {Barbone}}, \bibinfo {author} {\bibfnamefont {Pawel}\
  \bibnamefont {Latawiec}}, \bibinfo {author} {\bibfnamefont {Duhee}\
  \bibnamefont {Yoon}}, \bibinfo {author} {\bibfnamefont {Anna~K.}\
  \bibnamefont {Ott}}, \bibinfo {author} {\bibfnamefont {Marko}\ \bibnamefont
  {Loncar}}, \bibinfo {author} {\bibfnamefont {Andrea~C.}\ \bibnamefont
  {Ferrari}}, \ and\ \bibinfo {author} {\bibfnamefont {Mete}\ \bibnamefont
  {Atat{\"u}re}},\ }\bibfield  {title} {\enquote {\bibinfo {title}
  {{Large-scale quantum-emitter arrays in atomically thin semiconductors}},}\
  }\href {http://dx.doi.org/10.1038/ncomms15093} {\bibfield  {journal}
  {\bibinfo  {journal} {Nature Communications}\ }\textbf {\bibinfo {volume}
  {8}},\ \bibinfo {pages} {15093 EP --} (\bibinfo {year} {2017})},\ \bibinfo
  {note} {article}\BibitemShut {NoStop}%
\bibitem [{\citenamefont {Frindt}(1965)}]{Exfoliation_1965}%
  \BibitemOpen
  \bibfield  {author} {\bibinfo {author} {\bibfnamefont {R.~F.}\ \bibnamefont
  {Frindt}},\ }\bibfield  {title} {\enquote {\bibinfo {title} {{Optical
  Absorption of a Few Unit-Cell Layers of Mo${\mathrm{S}}_{2}$}},}\ }\href
  {\doibase 10.1103/PhysRev.140.A536} {\bibfield  {journal} {\bibinfo
  {journal} {Phys. Rev.}\ }\textbf {\bibinfo {volume} {140}},\ \bibinfo {pages}
  {A536--A539} (\bibinfo {year} {1965})}\BibitemShut {NoStop}%
\bibitem [{\citenamefont {Mennel}\ \emph {et~al.}(2018)\citenamefont {Mennel},
  \citenamefont {Furchi}, \citenamefont {Wachter}, \citenamefont {Paur},
  \citenamefont {Polyushkin},\ and\ \citenamefont
  {Mueller}}]{OpticalImagingOfStrain}%
  \BibitemOpen
  \bibfield  {author} {\bibinfo {author} {\bibfnamefont {Lukas}\ \bibnamefont
  {Mennel}}, \bibinfo {author} {\bibfnamefont {Marco~M.}\ \bibnamefont
  {Furchi}}, \bibinfo {author} {\bibfnamefont {Stefan}\ \bibnamefont
  {Wachter}}, \bibinfo {author} {\bibfnamefont {Matthias}\ \bibnamefont
  {Paur}}, \bibinfo {author} {\bibfnamefont {Dmitry~K.}\ \bibnamefont
  {Polyushkin}}, \ and\ \bibinfo {author} {\bibfnamefont {Thomas}\ \bibnamefont
  {Mueller}},\ }\bibfield  {title} {\enquote {\bibinfo {title} {{Optical
  imaging of strain in two-dimensional crystals}},}\ }\href {\doibase
  10.1038/s41467-018-02830-y} {\bibfield  {journal} {\bibinfo  {journal}
  {Nature Communications}\ }\textbf {\bibinfo {volume} {9}},\ \bibinfo {pages}
  {516} (\bibinfo {year} {2018})}\BibitemShut {NoStop}%
\bibitem [{\citenamefont {Hong}\ \emph {et~al.}(2015)\citenamefont {Hong},
  \citenamefont {Hu}, \citenamefont {Probert}, \citenamefont {Li},
  \citenamefont {Lv}, \citenamefont {Yang}, \citenamefont {Gu}, \citenamefont
  {Mao}, \citenamefont {Feng}, \citenamefont {Xie}, \citenamefont {Zhang},
  \citenamefont {Wu}, \citenamefont {Zhang}, \citenamefont {Jin}, \citenamefont
  {Ji}, \citenamefont {Zhang}, \citenamefont {Yuan},\ and\ \citenamefont
  {Zhang}}]{ExploringAtomicDefectsInMoS2}%
  \BibitemOpen
  \bibfield  {author} {\bibinfo {author} {\bibfnamefont {Jinhua}\ \bibnamefont
  {Hong}}, \bibinfo {author} {\bibfnamefont {Zhixin}\ \bibnamefont {Hu}},
  \bibinfo {author} {\bibfnamefont {Matt}\ \bibnamefont {Probert}}, \bibinfo
  {author} {\bibfnamefont {Kun}\ \bibnamefont {Li}}, \bibinfo {author}
  {\bibfnamefont {Danhui}\ \bibnamefont {Lv}}, \bibinfo {author} {\bibfnamefont
  {Xinan}\ \bibnamefont {Yang}}, \bibinfo {author} {\bibfnamefont {Lin}\
  \bibnamefont {Gu}}, \bibinfo {author} {\bibfnamefont {Nannan}\ \bibnamefont
  {Mao}}, \bibinfo {author} {\bibfnamefont {Qingliang}\ \bibnamefont {Feng}},
  \bibinfo {author} {\bibfnamefont {Liming}\ \bibnamefont {Xie}}, \bibinfo
  {author} {\bibfnamefont {Jin}\ \bibnamefont {Zhang}}, \bibinfo {author}
  {\bibfnamefont {Dianzhong}\ \bibnamefont {Wu}}, \bibinfo {author}
  {\bibfnamefont {Zhiyong}\ \bibnamefont {Zhang}}, \bibinfo {author}
  {\bibfnamefont {Chuanhong}\ \bibnamefont {Jin}}, \bibinfo {author}
  {\bibfnamefont {Wei}\ \bibnamefont {Ji}}, \bibinfo {author} {\bibfnamefont
  {Xixiang}\ \bibnamefont {Zhang}}, \bibinfo {author} {\bibfnamefont {Jun}\
  \bibnamefont {Yuan}}, \ and\ \bibinfo {author} {\bibfnamefont
  {Ze}~\bibnamefont {Zhang}},\ }\bibfield  {title} {\enquote {\bibinfo {title}
  {{Exploring atomic defects in molybdenum disulphide monolayers}},}\ }\href
  {http://dx.doi.org/10.1038/ncomms7293} {\bibfield  {journal} {\bibinfo
  {journal} {Nature Communications}\ }\textbf {\bibinfo {volume} {6}},\
  \bibinfo {pages} {6293 EP --} (\bibinfo {year} {2015})},\ \bibinfo {note}
  {article}\BibitemShut {NoStop}%
\bibitem [{\citenamefont {Wang}\ \emph {et~al.}(2015)\citenamefont {Wang},
  \citenamefont {Zhang},\ and\ \citenamefont
  {Rana}}]{UltrafastDynamicsofDefectAssistedRecombiniation}%
  \BibitemOpen
  \bibfield  {author} {\bibinfo {author} {\bibfnamefont {Haining}\ \bibnamefont
  {Wang}}, \bibinfo {author} {\bibfnamefont {Changjian}\ \bibnamefont {Zhang}},
  \ and\ \bibinfo {author} {\bibfnamefont {Farhan}\ \bibnamefont {Rana}},\
  }\bibfield  {title} {\enquote {\bibinfo {title} {{Ultrafast Dynamics of
  Defect-Assisted Electron-Hole Recombination in Monolayer
  $\mathrm{MoS_2}$}},}\ }\href {\doibase 10.1021/nl503636c} {\bibfield
  {journal} {\bibinfo  {journal} {Nano Letters}\ }\textbf {\bibinfo {volume}
  {15}},\ \bibinfo {pages} {339--345} (\bibinfo {year} {2015})}\BibitemShut
  {NoStop}%
\bibitem [{\citenamefont {Xue}\ \emph {et~al.}(2011)\citenamefont {Xue},
  \citenamefont {Sanchez-Yamagishi}, \citenamefont {Bulmash}, \citenamefont
  {Jacquod}, \citenamefont {Deshpande}, \citenamefont {Watanabe}, \citenamefont
  {Taniguchi}, \citenamefont {Jarillo-Herrero},\ and\ \citenamefont
  {LeRoy}}]{ScanningTunnelingMicroscopyOfGraphene}%
  \BibitemOpen
  \bibfield  {author} {\bibinfo {author} {\bibfnamefont {Jiamin}\ \bibnamefont
  {Xue}}, \bibinfo {author} {\bibfnamefont {Javier}\ \bibnamefont
  {Sanchez-Yamagishi}}, \bibinfo {author} {\bibfnamefont {Danny}\ \bibnamefont
  {Bulmash}}, \bibinfo {author} {\bibfnamefont {Philippe}\ \bibnamefont
  {Jacquod}}, \bibinfo {author} {\bibfnamefont {Aparna}\ \bibnamefont
  {Deshpande}}, \bibinfo {author} {\bibfnamefont {K.}~\bibnamefont {Watanabe}},
  \bibinfo {author} {\bibfnamefont {T.}~\bibnamefont {Taniguchi}}, \bibinfo
  {author} {\bibfnamefont {Pablo}\ \bibnamefont {Jarillo-Herrero}}, \ and\
  \bibinfo {author} {\bibfnamefont {Brian~J.}\ \bibnamefont {LeRoy}},\
  }\bibfield  {title} {\enquote {\bibinfo {title} {{Scanning tunnelling
  microscopy and spectroscopy of ultra-flat graphene on hexagonal boron
  nitride}},}\ }\href {http://dx.doi.org/10.1038/nmat2968} {\bibfield
  {journal} {\bibinfo  {journal} {Nature Materials}\ }\textbf {\bibinfo
  {volume} {10}},\ \bibinfo {pages} {282 EP --} (\bibinfo {year}
  {2011})}\BibitemShut {NoStop}%
\bibitem [{\citenamefont {Brechb\"uhler}\ \emph {et~al.}(2018)\citenamefont
  {Brechb\"uhler}, \citenamefont {Rabouw}, \citenamefont {Rohner},
  \citenamefont {le~Feber}, \citenamefont {Poulikakos},\ and\ \citenamefont
  {Norris}}]{PRL2018_SPP_Drexhage}%
  \BibitemOpen
  \bibfield  {author} {\bibinfo {author} {\bibfnamefont {Raphael}\ \bibnamefont
  {Brechb\"uhler}}, \bibinfo {author} {\bibfnamefont {Freddy~T.}\ \bibnamefont
  {Rabouw}}, \bibinfo {author} {\bibfnamefont {Patrik}\ \bibnamefont {Rohner}},
  \bibinfo {author} {\bibfnamefont {Boris}\ \bibnamefont {le~Feber}}, \bibinfo
  {author} {\bibfnamefont {Dimos}\ \bibnamefont {Poulikakos}}, \ and\ \bibinfo
  {author} {\bibfnamefont {David~J.}\ \bibnamefont {Norris}},\ }\bibfield
  {title} {\enquote {\bibinfo {title} {{Two-Dimensional Drexhage Experiment for
  Electric- and Magnetic-Dipole Sources on Plasmonic Interfaces}},}\ }\href
  {\doibase 10.1103/PhysRevLett.121.113601} {\bibfield  {journal} {\bibinfo
  {journal} {Phys. Rev. Lett.}\ }\textbf {\bibinfo {volume} {121}},\ \bibinfo
  {pages} {113601} (\bibinfo {year} {2018})}\BibitemShut {NoStop}%
\bibitem [{\citenamefont {Langguth}\ \emph {et~al.}(2016)\citenamefont
  {Langguth}, \citenamefont {Fleury}, \citenamefont {Al\`u},\ and\
  \citenamefont {Koenderink}}]{PRL_2016_Drexhage_Sound}%
  \BibitemOpen
  \bibfield  {author} {\bibinfo {author} {\bibfnamefont {Lutz}\ \bibnamefont
  {Langguth}}, \bibinfo {author} {\bibfnamefont {Romain}\ \bibnamefont
  {Fleury}}, \bibinfo {author} {\bibfnamefont {Andrea}\ \bibnamefont {Al\`u}},
  \ and\ \bibinfo {author} {\bibfnamefont {A.~Femius}\ \bibnamefont
  {Koenderink}},\ }\bibfield  {title} {\enquote {\bibinfo {title} {{Drexhage's
  Experiment for Sound}},}\ }\href {\doibase 10.1103/PhysRevLett.116.224301}
  {\bibfield  {journal} {\bibinfo  {journal} {Phys. Rev. Lett.}\ }\textbf
  {\bibinfo {volume} {116}},\ \bibinfo {pages} {224301} (\bibinfo {year}
  {2016})}\BibitemShut {NoStop}%
\bibitem [{\citenamefont {Wang}\ \emph {et~al.}(2016)\citenamefont {Wang},
  \citenamefont {Zhang}, \citenamefont {Chan}, \citenamefont {Manolatou},
  \citenamefont {Tiwari},\ and\ \citenamefont
  {Rana}}]{Farhan_PRB_2016_RadiativeLifetime}%
  \BibitemOpen
  \bibfield  {author} {\bibinfo {author} {\bibfnamefont {Haining}\ \bibnamefont
  {Wang}}, \bibinfo {author} {\bibfnamefont {Changjian}\ \bibnamefont {Zhang}},
  \bibinfo {author} {\bibfnamefont {Weimin}\ \bibnamefont {Chan}}, \bibinfo
  {author} {\bibfnamefont {Christina}\ \bibnamefont {Manolatou}}, \bibinfo
  {author} {\bibfnamefont {Sandip}\ \bibnamefont {Tiwari}}, \ and\ \bibinfo
  {author} {\bibfnamefont {Farhan}\ \bibnamefont {Rana}},\ }\bibfield  {title}
  {\enquote {\bibinfo {title} {{Radiative lifetimes of excitons and trions in
  monolayers of the metal dichalcogenide ${\mathrm{MoS}}_{2}$}},}\ }\href
  {\doibase 10.1103/PhysRevB.93.045407} {\bibfield  {journal} {\bibinfo
  {journal} {Phys. Rev. B}\ }\textbf {\bibinfo {volume} {93}},\ \bibinfo
  {pages} {045407} (\bibinfo {year} {2016})}\BibitemShut {NoStop}%
\bibitem [{\citenamefont {Soh}\ \emph {et~al.}(2018)\citenamefont {Soh},
  \citenamefont {Rogers}, \citenamefont {Gray}, \citenamefont {Chatterjee},\
  and\ \citenamefont {Mabuchi}}]{Daniel_MoS2_Nonlinearity}%
  \BibitemOpen
  \bibfield  {author} {\bibinfo {author} {\bibfnamefont {Daniel B.~S.}\
  \bibnamefont {Soh}}, \bibinfo {author} {\bibfnamefont {Christopher}\
  \bibnamefont {Rogers}}, \bibinfo {author} {\bibfnamefont {Dodd~J.}\
  \bibnamefont {Gray}}, \bibinfo {author} {\bibfnamefont {Eric}\ \bibnamefont
  {Chatterjee}}, \ and\ \bibinfo {author} {\bibfnamefont {Hideo}\ \bibnamefont
  {Mabuchi}},\ }\bibfield  {title} {\enquote {\bibinfo {title} {{Optical
  nonlinearities of excitons in monolayer ${\mathrm{MoS}}_{2}$}},}\ }\href
  {\doibase 10.1103/PhysRevB.97.165111} {\bibfield  {journal} {\bibinfo
  {journal} {Phys. Rev. B}\ }\textbf {\bibinfo {volume} {97}},\ \bibinfo
  {pages} {165111} (\bibinfo {year} {2018})}\BibitemShut {NoStop}%
\bibitem [{\citenamefont {Yu}\ \emph {et~al.}(2013)\citenamefont {Yu},
  \citenamefont {Li}, \citenamefont {Liu}, \citenamefont {Su}, \citenamefont
  {Zhang},\ and\ \citenamefont
  {Cao}}]{ScalableSynthesisofUniformMonolayerMoS2}%
  \BibitemOpen
  \bibfield  {author} {\bibinfo {author} {\bibfnamefont {Yifei}\ \bibnamefont
  {Yu}}, \bibinfo {author} {\bibfnamefont {Chun}\ \bibnamefont {Li}}, \bibinfo
  {author} {\bibfnamefont {Yi}~\bibnamefont {Liu}}, \bibinfo {author}
  {\bibfnamefont {Liqin}\ \bibnamefont {Su}}, \bibinfo {author} {\bibfnamefont
  {Yong}\ \bibnamefont {Zhang}}, \ and\ \bibinfo {author} {\bibfnamefont
  {Linyou}\ \bibnamefont {Cao}},\ }\bibfield  {title} {\enquote {\bibinfo
  {title} {{Controlled Scalable Synthesis of Uniform, High-Quality Monolayer
  and Few-layer $\mathrm{MoS_2}$ Films}},}\ }\href
  {http://dx.doi.org/10.1038/srep01866} {\bibfield  {journal} {\bibinfo
  {journal} {Scientific Reports}\ }\textbf {\bibinfo {volume} {3}},\ \bibinfo
  {pages} {1866 EP --} (\bibinfo {year} {2013})},\ \bibinfo {note}
  {article}\BibitemShut {NoStop}%
\bibitem [{\citenamefont {Lee}\ \emph {et~al.}()\citenamefont {Lee},
  \citenamefont {Zhang}, \citenamefont {Zhang}, \citenamefont {Chang},
  \citenamefont {Lin}, \citenamefont {Chang}, \citenamefont {Yu}, \citenamefont
  {Wang}, \citenamefont {Chang}, \citenamefont {Li},\ and\ \citenamefont
  {Lin}}]{SynthesisOfLargeAreaMoS2}%
  \BibitemOpen
  \bibfield  {author} {\bibinfo {author} {\bibfnamefont {Yi‐Hsien}\
  \bibnamefont {Lee}}, \bibinfo {author} {\bibfnamefont {Xin‐Quan}\
  \bibnamefont {Zhang}}, \bibinfo {author} {\bibfnamefont {Wenjing}\
  \bibnamefont {Zhang}}, \bibinfo {author} {\bibfnamefont {Mu‐Tung}\
  \bibnamefont {Chang}}, \bibinfo {author} {\bibfnamefont {Cheng‐Te}\
  \bibnamefont {Lin}}, \bibinfo {author} {\bibfnamefont {Kai‐Di}\
  \bibnamefont {Chang}}, \bibinfo {author} {\bibfnamefont {Ya‐Chu}\
  \bibnamefont {Yu}}, \bibinfo {author} {\bibfnamefont {Jacob~Tse‐Wei}\
  \bibnamefont {Wang}}, \bibinfo {author} {\bibfnamefont {Chia‐Seng}\
  \bibnamefont {Chang}}, \bibinfo {author} {\bibfnamefont {Lain‐Jong}\
  \bibnamefont {Li}}, \ and\ \bibinfo {author} {\bibfnamefont {Tsung‐Wu}\
  \bibnamefont {Lin}},\ }\bibfield  {title} {\enquote {\bibinfo {title}
  {{Synthesis of Large‐Area $\mathrm{MoS_2}$ Atomic Layers with Chemical
  Vapor Deposition}},}\ }\href {\doibase 10.1002/adma.201104798} {\bibfield
  {journal} {\bibinfo  {journal} {Advanced Materials}\ }\textbf {\bibinfo
  {volume} {24}},\ \bibinfo {pages} {2320--2325}}\BibitemShut {NoStop}%
\bibitem [{\citenamefont {Rogers}\ \emph {et~al.}(2018)\citenamefont {Rogers},
  \citenamefont {Gray}, \citenamefont {Bogdanowicz},\ and\ \citenamefont
  {Mabuchi}}]{LaserAnnealing}%
  \BibitemOpen
  \bibfield  {author} {\bibinfo {author} {\bibfnamefont {Christopher}\
  \bibnamefont {Rogers}}, \bibinfo {author} {\bibfnamefont {Dodd}\ \bibnamefont
  {Gray}}, \bibinfo {author} {\bibfnamefont {Nate}\ \bibnamefont
  {Bogdanowicz}}, \ and\ \bibinfo {author} {\bibfnamefont {Hideo}\ \bibnamefont
  {Mabuchi}},\ }\bibfield  {title} {\enquote {\bibinfo {title} {{Laser
  annealing for radiatively broadened ${\mathrm{MoSe}}_{2}$ grown by chemical
  vapor deposition}},}\ }\href {\doibase 10.1103/PhysRevMaterials.2.094003}
  {\bibfield  {journal} {\bibinfo  {journal} {Phys. Rev. Materials}\ }\textbf
  {\bibinfo {volume} {2}},\ \bibinfo {pages} {094003} (\bibinfo {year}
  {2018})}\BibitemShut {NoStop}%
\bibitem [{\citenamefont {Zomer}\ \emph {et~al.}(2014)\citenamefont {Zomer},
  \citenamefont {Guimarães}, \citenamefont {Brant}, \citenamefont {Tombros},\
  and\ \citenamefont {van Wees}}]{FastPickupTechnique}%
  \BibitemOpen
  \bibfield  {author} {\bibinfo {author} {\bibfnamefont {P.~J.}\ \bibnamefont
  {Zomer}}, \bibinfo {author} {\bibfnamefont {M.~H.~D.}\ \bibnamefont
  {Guimarães}}, \bibinfo {author} {\bibfnamefont {J.~C.}\ \bibnamefont
  {Brant}}, \bibinfo {author} {\bibfnamefont {N.}~\bibnamefont {Tombros}}, \
  and\ \bibinfo {author} {\bibfnamefont {B.~J.}\ \bibnamefont {van Wees}},\
  }\bibfield  {title} {\enquote {\bibinfo {title} {{Fast pick up technique for
  high quality heterostructures of bilayer graphene and hexagonal boron
  nitride}},}\ }\href {\doibase 10.1063/1.4886096} {\bibfield  {journal}
  {\bibinfo  {journal} {Applied Physics Letters}\ }\textbf {\bibinfo {volume}
  {105}},\ \bibinfo {pages} {013101} (\bibinfo {year} {2014})}\BibitemShut
  {NoStop}%
\bibitem [{\citenamefont {Pizzocchero}\ \emph {et~al.}(2016)\citenamefont
  {Pizzocchero}, \citenamefont {Gammelgaard}, \citenamefont {Jessen},
  \citenamefont {Caridad}, \citenamefont {Wang}, \citenamefont {Hone},
  \citenamefont {B{\o}ggild},\ and\ \citenamefont
  {Booth}}]{HotPickupTechnique}%
  \BibitemOpen
  \bibfield  {author} {\bibinfo {author} {\bibfnamefont {Filippo}\ \bibnamefont
  {Pizzocchero}}, \bibinfo {author} {\bibfnamefont {Lene}\ \bibnamefont
  {Gammelgaard}}, \bibinfo {author} {\bibfnamefont {Bjarke~S.}\ \bibnamefont
  {Jessen}}, \bibinfo {author} {\bibfnamefont {Jos{\'e}~M.}\ \bibnamefont
  {Caridad}}, \bibinfo {author} {\bibfnamefont {Lei}\ \bibnamefont {Wang}},
  \bibinfo {author} {\bibfnamefont {James}\ \bibnamefont {Hone}}, \bibinfo
  {author} {\bibfnamefont {Peter}\ \bibnamefont {B{\o}ggild}}, \ and\ \bibinfo
  {author} {\bibfnamefont {Timothy~J.}\ \bibnamefont {Booth}},\ }\bibfield
  {title} {\enquote {\bibinfo {title} {{The hot pick-up technique for batch
  assembly of van der Waals heterostructures}},}\ }\href
  {https://doi.org/10.1038/ncomms11894} {\bibfield  {journal} {\bibinfo
  {journal} {Nature Communications}\ }\textbf {\bibinfo {volume} {7}},\
  \bibinfo {pages} {11894 EP --} (\bibinfo {year} {2016})},\ \bibinfo {note}
  {article}\BibitemShut {NoStop}%
\bibitem [{\citenamefont {Wooten}(1972)}]{WootenOpticalPropertiesOfSolids}%
  \BibitemOpen
  \bibfield  {author} {\bibinfo {author} {\bibfnamefont {F.}~\bibnamefont
  {Wooten}},\ }\href {https://books.google.com/books?id=A\_dHNRXFq28C} {\emph
  {\bibinfo {title} {{Optical properties of solids}}}}\ (\bibinfo  {publisher}
  {Academic Press},\ \bibinfo {year} {1972})\ Chap.~\bibinfo {chapter}
  {3}\BibitemShut {NoStop}%
\bibitem [{\citenamefont {Katsidis}\ and\ \citenamefont
  {Siapkas}(2002)}]{TranferMatrixMethod}%
  \BibitemOpen
  \bibfield  {author} {\bibinfo {author} {\bibfnamefont {Charalambos~C.}\
  \bibnamefont {Katsidis}}\ and\ \bibinfo {author} {\bibfnamefont
  {Dimitrios~I.}\ \bibnamefont {Siapkas}},\ }\bibfield  {title} {\enquote
  {\bibinfo {title} {{General transfer-matrix method for optical multilayer
  systems with coherent, partially coherent, and incoherent interference}},}\
  }\href {\doibase 10.1364/AO.41.003978} {\bibfield  {journal} {\bibinfo
  {journal} {Appl. Opt.}\ }\textbf {\bibinfo {volume} {41}},\ \bibinfo {pages}
  {3978--3987} (\bibinfo {year} {2002})}\BibitemShut {NoStop}%
\bibitem [{\citenamefont {Liu}\ \emph {et~al.}(2001)\citenamefont {Liu},
  \citenamefont {Lin}, \citenamefont {Huang}, \citenamefont {Guo},\ and\
  \citenamefont {Duan}}]{EmpiricalApproximationVoight}%
  \BibitemOpen
  \bibfield  {author} {\bibinfo {author} {\bibfnamefont {Yuyan}\ \bibnamefont
  {Liu}}, \bibinfo {author} {\bibfnamefont {Jieli}\ \bibnamefont {Lin}},
  \bibinfo {author} {\bibfnamefont {Guangming}\ \bibnamefont {Huang}}, \bibinfo
  {author} {\bibfnamefont {Yuanqing}\ \bibnamefont {Guo}}, \ and\ \bibinfo
  {author} {\bibfnamefont {Chuanxi}\ \bibnamefont {Duan}},\ }\bibfield  {title}
  {\enquote {\bibinfo {title} {{Simple empirical analytical approximation to
  the Voigt profile}},}\ }\href {\doibase 10.1364/JOSAB.18.000666} {\bibfield
  {journal} {\bibinfo  {journal} {J. Opt. Soc. Am. B}\ }\textbf {\bibinfo
  {volume} {18}},\ \bibinfo {pages} {666--672} (\bibinfo {year}
  {2001})}\BibitemShut {NoStop}%
\bibitem [{\citenamefont {Olivero}\ and\ \citenamefont
  {Longbothum}(1977)}]{VoightBriefReview}%
  \BibitemOpen
  \bibfield  {author} {\bibinfo {author} {\bibfnamefont {J.J.}\ \bibnamefont
  {Olivero}}\ and\ \bibinfo {author} {\bibfnamefont {R.L.}\ \bibnamefont
  {Longbothum}},\ }\bibfield  {title} {\enquote {\bibinfo {title} {{Empirical
  fits to the Voigt line width: A brief review}},}\ }\href {\doibase
  https://doi.org/10.1016/0022-4073(77)90161-3} {\bibfield  {journal} {\bibinfo
   {journal} {Journal of Quantitative Spectroscopy and Radiative Transfer}\
  }\textbf {\bibinfo {volume} {17}},\ \bibinfo {pages} {233 -- 236} (\bibinfo
  {year} {1977})}\BibitemShut {NoStop}%
\bibitem [{\citenamefont {{Zhou}}\ \emph {et~al.}(2019)\citenamefont {{Zhou}},
  \citenamefont {{Scuri}}, \citenamefont {{Sung}}, \citenamefont {{Gelly}},
  \citenamefont {{Wild}}, \citenamefont {{De Greve}}, \citenamefont {{Joe}},
  \citenamefont {{Taniguchi}}, \citenamefont {{Watanabe}}, \citenamefont
  {{Kim}}, \citenamefont {{Lukin}},\ and\ \citenamefont
  {{Park}}}]{ControllingExcitonsWithAMirror}%
  \BibitemOpen
  \bibfield  {author} {\bibinfo {author} {\bibfnamefont {You}\ \bibnamefont
  {{Zhou}}}, \bibinfo {author} {\bibfnamefont {Giovanni}\ \bibnamefont
  {{Scuri}}}, \bibinfo {author} {\bibfnamefont {Jiho}\ \bibnamefont {{Sung}}},
  \bibinfo {author} {\bibfnamefont {Ryan~J.}\ \bibnamefont {{Gelly}}}, \bibinfo
  {author} {\bibfnamefont {Dominik~S.}\ \bibnamefont {{Wild}}}, \bibinfo
  {author} {\bibfnamefont {Kristiaan}\ \bibnamefont {{De Greve}}}, \bibinfo
  {author} {\bibfnamefont {Andrew~Y.}\ \bibnamefont {{Joe}}}, \bibinfo {author}
  {\bibfnamefont {Takashi}\ \bibnamefont {{Taniguchi}}}, \bibinfo {author}
  {\bibfnamefont {Kenji}\ \bibnamefont {{Watanabe}}}, \bibinfo {author}
  {\bibfnamefont {Philip}\ \bibnamefont {{Kim}}}, \bibinfo {author}
  {\bibfnamefont {Mikhail~D.}\ \bibnamefont {{Lukin}}}, \ and\ \bibinfo
  {author} {\bibfnamefont {Hongkun}\ \bibnamefont {{Park}}},\ }\bibfield
  {title} {\enquote {\bibinfo {title} {{Controlling excitons in an atomically
  thin membrane with a mirror}},}\ }\href@noop {} {\bibfield  {journal}
  {\bibinfo  {journal} {arXiv e-prints}\ ,\ \bibinfo {eid} {arXiv:1901.08500}}
  (\bibinfo {year} {2019})},\ \Eprint {http://arxiv.org/abs/1901.08500}
  {arXiv:1901.08500 [cond-mat.mes-hall]} \BibitemShut {NoStop}%
\bibitem [{\citenamefont {Bogdanowicz}\ \emph {et~al.}(2018)\citenamefont
  {Bogdanowicz}, \citenamefont {Rogers}, \citenamefont {Gray}, \citenamefont
  {Timossi}, \citenamefont {Marazzi}, \citenamefont {Wheeler},\ and\
  \citenamefont {Galinskiy}}]{Instrumental}%
  \BibitemOpen
  \bibfield  {author} {\bibinfo {author} {\bibfnamefont {Nate}\ \bibnamefont
  {Bogdanowicz}}, \bibinfo {author} {\bibfnamefont {Christopher}\ \bibnamefont
  {Rogers}}, \bibinfo {author} {\bibfnamefont {Dodd}\ \bibnamefont {Gray}},
  \bibinfo {author} {\bibfnamefont {Chris}\ \bibnamefont {Timossi}}, \bibinfo
  {author} {\bibfnamefont {Francesco}\ \bibnamefont {Marazzi}}, \bibinfo
  {author} {\bibfnamefont {Jonathan}\ \bibnamefont {Wheeler}}, \ and\ \bibinfo
  {author} {\bibfnamefont {Ivan}\ \bibnamefont {Galinskiy}},\ }\bibfield
  {title} {\enquote {\bibinfo {title} {{mabuchilab/Instrumental: 0.5}},}\
  }\href {\doibase 10.5281/zenodo.2556399} {\  (\bibinfo {year} {2018}),\
  10.5281/zenodo.2556399}\BibitemShut {NoStop}%
\end{thebibliography}%


\begin{thebibliography}{0}%
\makeatletter
\providecommand \@ifxundefined [1]{%
 \@ifx{#1\undefined}
}%
\providecommand \@ifnum [1]{%
 \ifnum #1\expandafter \@firstoftwo
 \else \expandafter \@secondoftwo
 \fi
}%
\providecommand \@ifx [1]{%
 \ifx #1\expandafter \@firstoftwo
 \else \expandafter \@secondoftwo
 \fi
}%
\providecommand \natexlab [1]{#1}%
\providecommand \enquote  [1]{``#1''}%
\providecommand \bibnamefont  [1]{#1}%
\providecommand \bibfnamefont [1]{#1}%
\providecommand \citenamefont [1]{#1}%
\providecommand \href@noop [0]{\@secondoftwo}%
\providecommand \href [0]{\begingroup \@sanitize@url \@href}%
\providecommand \@href[1]{\@@startlink{#1}\@@href}%
\providecommand \@@href[1]{\endgroup#1\@@endlink}%
\providecommand \@sanitize@url [0]{\catcode `\\12\catcode `\$12\catcode
  `\&12\catcode `\#12\catcode `\^12\catcode `\_12\catcode `\%12\relax}%
\providecommand \@@startlink[1]{}%
\providecommand \@@endlink[0]{}%
\providecommand \url  [0]{\begingroup\@sanitize@url \@url }%
\providecommand \@url [1]{\endgroup\@href {#1}{\urlprefix }}%
\providecommand \urlprefix  [0]{URL }%
\providecommand \Eprint [0]{\href }%
\providecommand \doibase [0]{http://dx.doi.org/}%
\providecommand \selectlanguage [0]{\@gobble}%
\providecommand \bibinfo  [0]{\@secondoftwo}%
\providecommand \bibfield  [0]{\@secondoftwo}%
\providecommand \translation [1]{[#1]}%
\providecommand \BibitemOpen [0]{}%
\providecommand \bibitemStop [0]{}%
\providecommand \bibitemNoStop [0]{.\EOS\space}%
\providecommand \EOS [0]{\spacefactor3000\relax}%
\providecommand \BibitemShut  [1]{\csname bibitem#1\endcsname}%
\let\auto@bib@innerbib\@empty
\end{thebibliography}%
\newpage

\section{\label{sec:SI} Additional Information}
Supplementary Information is linked to the online version of the paper.

\begin{acknowledgments}
This work was funded in part by the National Science Foundation (NSF) awards PHY-1648807 and DMR-1838497, and also by a seed grant from the Precourt Institute for Energy at Stanford University.
Part of this work was performed at the Stanford Nano Shared Facilities (SNSF), supported by the NSF under award ECCS-1542152.
C.R., D.G., and N.B. were supported in part by Stanford Graduate Fellowships.
C.R. was also supported in part by a Natural Sciences and Engineering Research Council of Canada doctoral postgraduate scholarship.
K.W. and T.T. acknowledge support from the Elemental Strategy Initiative conducted by the MEXT, Japan and and the CREST (JPMJCR15F3), JST.
All authors thank Tatsuhiro Onodera for discussions relating to the reflectance model.
All authors thank Daniel B. S. Soh and Eric Chatterjee for discussions relating to exciton physics.
C.R. thanks Joe Finney, Giovanni Scuri and Elyse Barre for help with heterostructure fabrication, Logan Wright for pointing out relevant literature, and Peter L. McMahon for suggestions relating to the manuscript.
\end{acknowledgments}

\section{\label{sec:Contributions} Author Contributions}
C.R. conceived the experiments, fabricated the samples, performed the experiments, and performed the data analysis.
C.R. and D.G. built the confocal microscope setup for the cryostat, as well as the grating spectrometer.
C.R. and N.B. automated the measurements.
T.T. and K.W. grew the hBN.
C.R., N.B., D.G. and H.M. all contributed to the manuscript.

\section{\label{sec:FinancialInterests} Competing financial interest}
The authors declare no competing financial interests.

\section{\label{sec:Data Availability} Data availability}
The data that support the findings of this study are available from the corresponding author upon reasonable request.

\end{document}




\title{Coherent Control of Two-Dimensional Excitons \\ Supplemental Methods}

\author{Christopher Rogers}
\email{cmrogers@stanford.edu}
\affiliation{%
 Ginzton Laboratory, Stanford University, 348 Via Pueblo, Stanford, CA 94305
}%
\author{Dodd Gray, Jr.}
\affiliation{%
 Ginzton Laboratory, Stanford University, 348 Via Pueblo, Stanford, CA 94305
}%
\author{Nathan Bogdanowicz}
\affiliation{%
 Ginzton Laboratory, Stanford University, 348 Via Pueblo, Stanford, CA 94305
}%
\author{Takashi Taniguchi}
\affiliation{%
National Institute for Materials Science, 1-1 Namiki, Tsukuba 305-0044, Japan
}%
\author{Kenji Watanabe}
\affiliation{%
National Institute for Materials Science, 1-1 Namiki, Tsukuba 305-0044, Japan
}%
\author{Hideo Mabuchi}
\email{hmabuchi@stanford.edu}
\affiliation{%
 Ginzton Laboratory, Stanford University, 348 Via Pueblo, Stanford, CA 94305
}%
%
%
%

\date{\today}
\maketitle


\beginsupplement


\section{\label{sec:Supplement} Supplemental Methods}
\subsection{\label{subsec:Supplement:Samples} Sample Fabrication}
We fabricate heterostructures using a dry pickup transfer technique [42, 43].
We first clean 300 nm $\mathrm{SiO_2}$ on Si substrates, and fused silica substrates by sonicating in acetone for 2 minutes, then deionized water for 2 minutes and finally isopropanol for 2 minutes.
The substrates are then subjected to oxygen plasma for 5 minutes.
Graphite (NGS Naturegraphit GmbH), hexagonal Boron Nitride (hBN), and $\mathrm{MoSe_2}$ (2D Semiconductors or HQ Graphene) are then exfoliated onto the freshly cleaned substrates using Scotch tape.
The substrates are observed under an optical microscope to identify monolayer $\mathrm{MoSe_2}$, few-layer graphene and 50-120 nm hBN.

Polydimethylsiloxane (PDMS) with thin polycarbonate (PC) stamps are used to create the heterostructures.
To produce the stamp, a $6 \%$ PC solution is used to form a thin film on a glass slide.
This thin film is then transferred onto a 1 mm $\times$ 1 mm piece of PDMS on a different glass slide using Scotch tape with a hole punched in the middle.
This stamp is then used to sequentially pick up the mechanically exfoliated flakes by bringing the stamp slowly into contact with a flake on the exfoliation substrate.
In our case, we first pick up the `top' hBN, then the monolayer $\mathrm{MoSe_2}$, then the `bottom' hBN, and finally the few-layer graphene flake.
Each flake is picked up at a temperature of about \SI{60}{\celsius}.
This stack (including the PC film) is transferred to a glass substrate by heating the substrate to 140 $^\circ$C and bringing the stamp into contact.
After letting the sample sit for one day, the PC is removed by dissolution in chloroform.

\subsection{\label{subsec:Supplement:Setup} Experimental Setup}
\begin{figure}
    \centering
    \subfloat[]
    {\includegraphics[width=1.\textwidth]
    {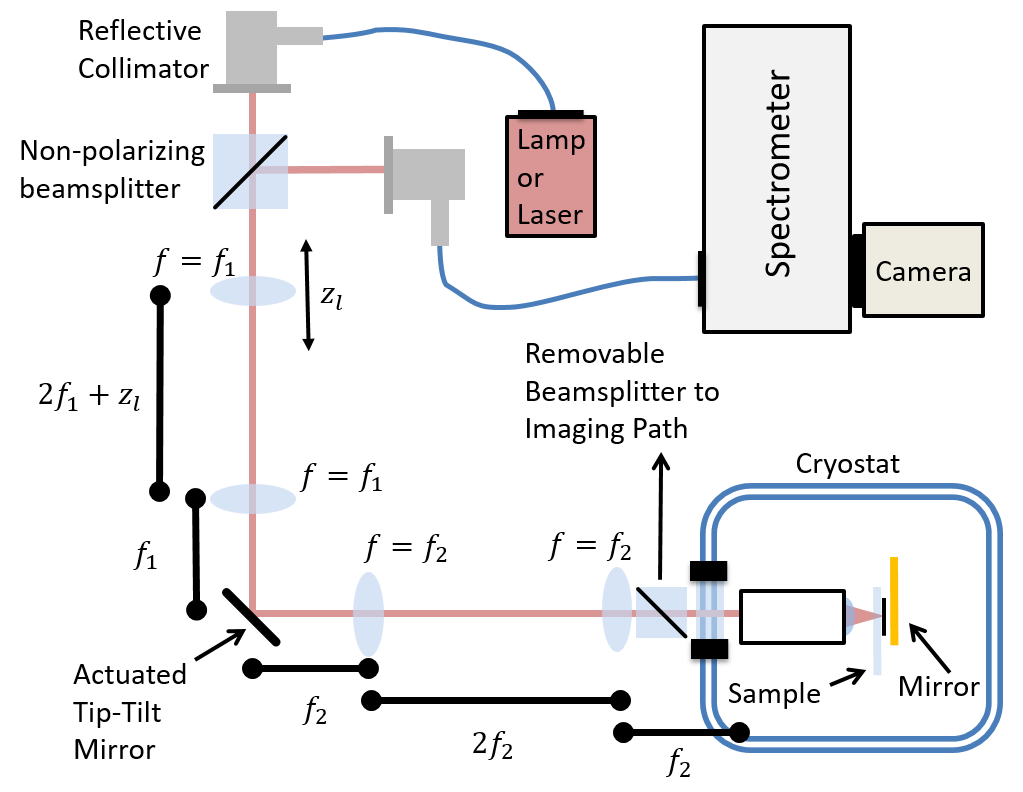}}
    \caption{\textbf{Experimental Setup.} Light is coupled from a lamp or a laser through a single mode fiber into a custom confocal microscope, which focuses light on the sample and collects the reflection.
    The collected light is measured using a grating spectrometer.
    \label{sfig:ExperimentSchematic}}
\end{figure}

A detailed experimental schematic is shown in Fig. \ref{sfig:ExperimentSchematic}.
The experiment is conducted in an optical cryostat (Montana Instruments Nanoscale Workstation) at a nominal temperature of 4 K and a pressure of $1\cdot 10^{-7}$ Torr.
The sample is attached to a fixed mount while the gold mirror is actuated by a slip-stick piezo mirror mount (Janssen Precision Engineering).

Light from either a lamp (Thorlabs SLS201) or a supercontinuum laser (NKT Photonics SuperK) is coupled into the custom confocal microscope through a single mode optical fiber.
Two reflective collimators serve to couple the microscope to the single mode fibers for excitation and reflection.
The excitation and reflection paths are separated by a 50/50 non-polarizing beamsplitter.
Two achromatic lenses with focal length $f_1=75$mm form the first 4f system.
The first of these lenses is translated along the optical axis using a motorized stage, which shifts the focus of the beam at the sample along the optical axis.
The range of travel of the beam focus at the sample is approximately $\pm 200$ \si{\micro\metre}.
A tip-tilt mirror mechanically actuated by motorized stages (Newport U100-A and Newport LTA-HS) at the beginning of the second 4f section (comprised of two achromatic lenses of focal length $f_2 = 150$ mm) shifts the beam in the transverse plane at the sample.
The total travel of the beam focus is about $\pm 300$ \si{\micro\metre}.
A microscope objective ($20\times$, 0.4 numerical aperture, Olympus MSPLAN) inside the optical cryostat focuses the light down on the sample and mirror.
Light is collected back through the same optical path, and sent to a grating spectrometer for measurement.
A removable beamsplitter enables imaging of the sample.
Note that the lens imaging onto the camera in the imaging train is also on a translation stage, allowing the imaging plane to be matched with that of the excitation spot.

The grating spectrometer used to measure the reflectance spectra has an 1800 line/mm reflective diffraction grating on a motorized rotation stage (Newport RGV100).
Spectra are measured using a camera (Princeton Instruments PIXIS 2048).
The nominal resolution of the spectrometer is approximately 1 \si{\cm^{-1}}.

\subsection{\label{subsec:Supplement:GlobalFits} Reflectance Model Fitting}
\begin{figure}
    \centering

    \subfloat[]
    {\includegraphics[width=0.9\textwidth]
    {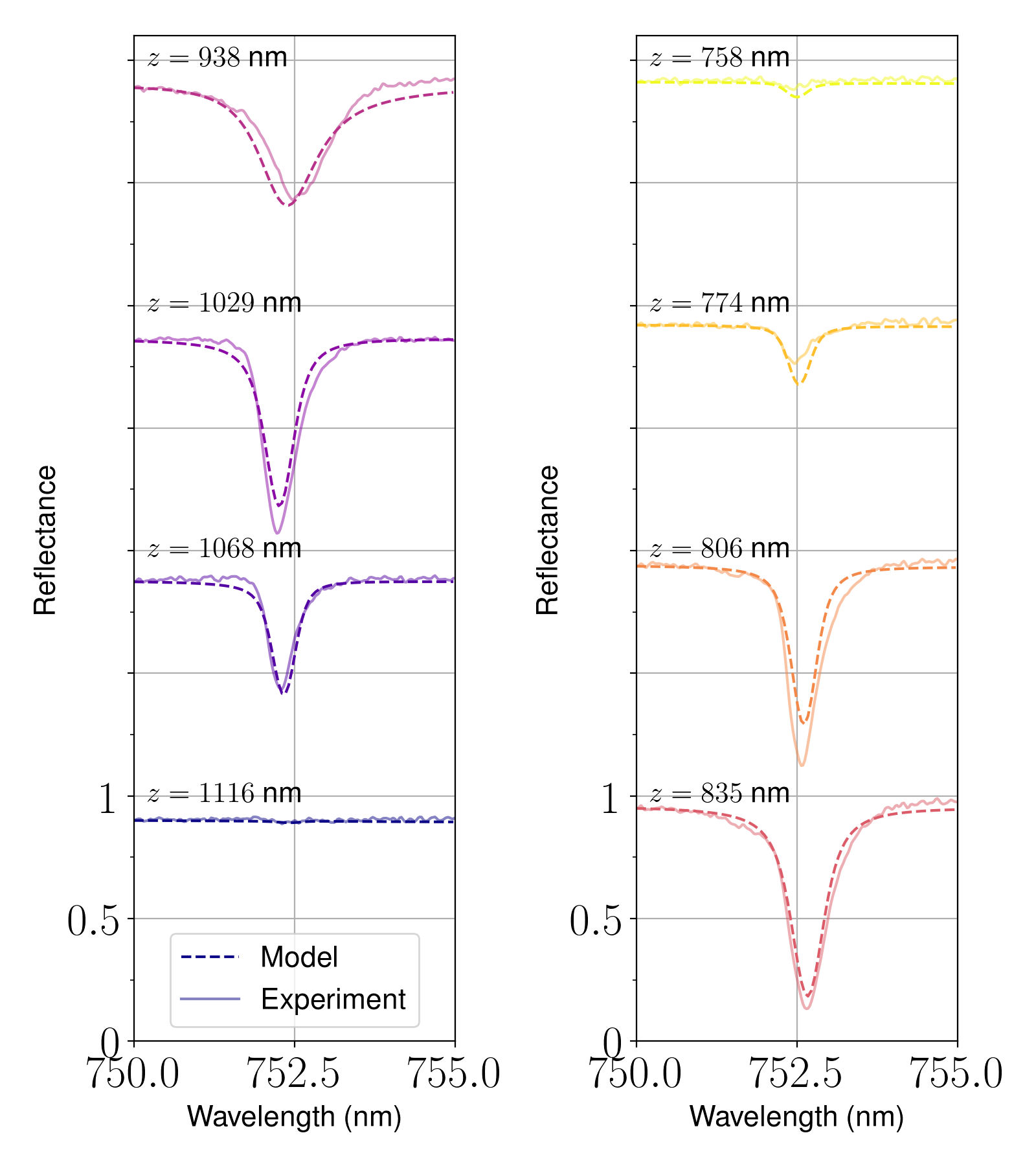}}
    \caption{\textbf{Characteristic spectra used to fit the model parameters.}
    The experimental data is shown in solid lines, and the model in dashed lines.
    The mirror position $z$ corresponding to each spectrum is shown.
    \label{sfig:FitSpectra}
    }
\end{figure}

We simultaneously fit spectral data in Figs. 2a and 2c from several selected characteristic $z$ positions to the full reflectance model in Eq. 3.
The absolute value of the discrepancy between the reflectance model and the experimental data over the selected $z$ values was simultaneously minimized to find the global fitting parameters $\omega_0=1647.72$ meV, $\gamma_{r, 0} = 1.09$ meV, $\gamma_{nr}=0.40$ meV and $\gamma_{ib}=0.26$ meV.
The experimental spectra selected for the fitting procedure and the corresponding modeled reflectance given by the optimized fitting parameters are shown in Fig. \ref{sfig:FitSpectra}.

We note that the four fitting parameters are highly constrained by the experimental data, and that other values of the parameters do not produce satisfactory agreement between experiment and model.
For simplicity we ignore the subtle difference between $\gamma_{ib}$ and $\gamma_{ib, \mathrm{eff}}$.
Doing so does not qualitatively alter the conclusions reached.
First, $\omega_0$ is set by the position of the reflectance dip at both $z_c$ and $z_{d, 1/2}$.
Second, the quantity $\gamma_{nr} + \gamma_{ib}$ is constrained by the reflectance linewidth at $z_d$, where radiative broadening is negligible.
Similarly, the total linewidth $\gamma_\mathrm{tot} = \gamma_r + \gamma_{nr} + \gamma_{ib}$ is constrained by the linewidth at $z_c$.
Lastly, the magnitude of the on-resonant reflection at $z_c$ constrains the ratio $\gamma_r/(\gamma_{nr} + \gamma_{ib})$.
Because the modulation of $\gamma_r$ by the mirror is independent of the fitting parameters, we can conceptually replace $\gamma_r$ in the above discussion by $A\gamma_{r,0}$ (where $A$ is constant).
The four independent relations above then fully constrain the fitting parameters.

The static (unfitted) parameters used in the reflectance model are as follows.
The index of the silica substrate is $n =1.45$, and the index of the hBN is $n=1.9$.
The index of refraction of the gold at $\omega_0$ is $n = 0.1388 + 4.4909i$.
The thickness of the gold is 120 nm.
The thickness of the top hBN is 87 nm, and the thickness of the bottom hBN is 128 nm.
The background index of the $\mathrm{MoSe_2}$ is $n = 4.5$.
The graphene flake is modeled as a bilayer with an index at $\omega_0$ of $n = 2.15 + 1.91i$.

\subsection{\label{subsec:Supplement:ZFits} Mirror Position Fitting}
\begin{figure}
    \centering

    \subfloat[]
    {\includegraphics[width=0.8\textwidth]
    {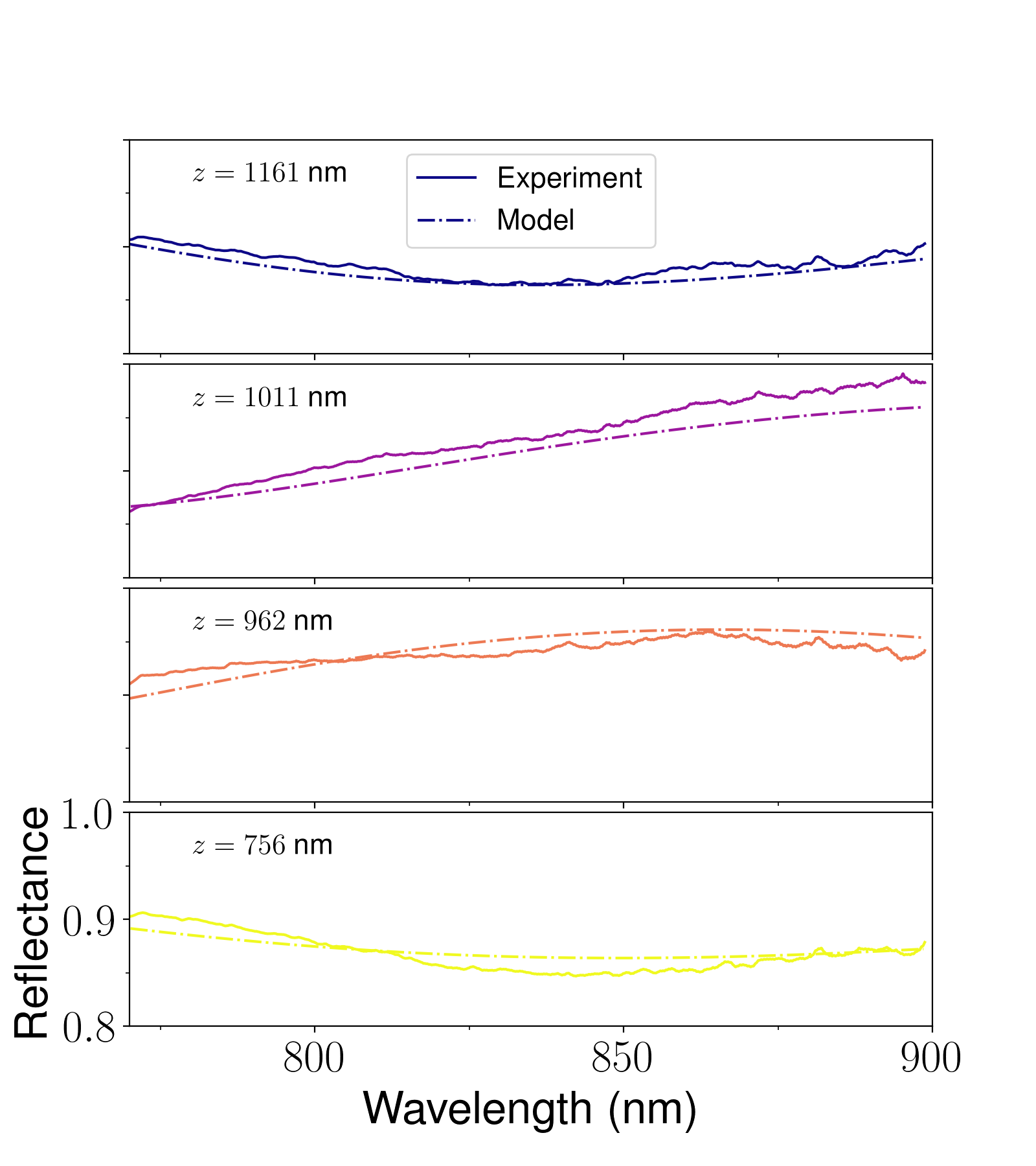}}
    \caption{\textbf{Selected spectra used to extract the mirror position $\boldsymbol{z}$.}
    Data from the experiment is shown with solid lines, and the model is shown with dashed lines.
    The mirror position $z$ corresponding to each spectrum is labeled.
    \label{sfig:ZFitSpectra}
    }
\end{figure}

The slip-stick piezo stage used to actuate the mirror does not have any position encoding, which requires us to have an independent measure of the mirror position.
At each $z$ position for the data in Figs. 2a and 2c, we also took spectra over the range of 770 nm to 900 nm from the same position on the sample.
As can be seen in Fig. \ref{sfig:ZFitSpectra}, there are broad fringes that vary with mirror position, due primarily to the modulation of absorption in the gold mirror and the few-layer graphene as the mirror position is changed.
Treating $z$ as a free parameter, we fit the reflectance in this region to the same reflectance model for the full heterostructure shown in Eq. 3.
Note that because these spectra are off-resonant from $X_0$, the exciton susceptibility has a negligible effect.
Several examples of the measured and fitted spectra are shown in Fig. \ref{sfig:ZFitSpectra}.

\subsection{\label{subsec:Supplement:Linewidths} Linewidth Model}
From [5], for an ideal dipole near and parallel to an ideal mirror:
\begin{equation}
\frac{\tau_x}{\tau_0} = \left[ 1-\frac{3\sin x}{2x} - \frac{3\cos x}{2x^2} + \frac{3\sin x}{2x} \right]^{-1}
\label{SIeq:drexhage_tau}
\end{equation}
where $\tau_x$ is the lifetime at normalized distance $x=\frac{4 \pi z}{\lambda_0}$ from the mirror, $\tau_0$ is the lifetime in vacuum, $\lambda_0$ is the wavelength in vacuum and $z$ is the {\em optical} path length between the mirror and the dipole.
When the dipole has a coherent quantum efficiency $\eta_0$ in vacuum, the modified lifetime $\tau^\prime_x$ is:
\begin{equation}
\frac{\tau^\prime_x}{\tau_0} = \frac{1}{1 + \eta_0 \left(\frac{\tau_0}{\tau_x} - 1\right)}
\label{SIeq:drexhage_tau_eta}
\end{equation}
It then follows from Eq. \ref{SIeq:drexhage_tau} that the radiative decay rate $\gamma_r$ for a perfect dipole is:
\begin{equation}
\frac{\gamma_r}{\gamma_{r,0}} = 1-\frac{3\sin x}{2x} - \frac{3\cos x}{2x^2} + \frac{3\sin x}{2x^3}
\label{SIeq:drexhage_gamma}
\end{equation}
where $\gamma_{r,0}$ is the radiative decay rate in vacuum.
For the more general case with sub-unity coherent quantum efficiency $\eta_0=\frac{\gamma_{r, 0}}{\gamma_{\mathrm{tot}, 0}}$, with $\gamma_{\mathrm{tot}, 0}$ being the total linewidth in vacuum, it follows from Eq. \ref{SIeq:drexhage_tau_eta} that:
\begin{equation}
\frac{\gamma_{\mathrm{tot}}(x)}{\gamma_{\mathrm{tot}, 0}} = 1 + \eta_0 \left(\frac{\gamma_r}{\gamma_{r,0}} - 1\right)
\label{SIeq:drexhage_gamma_eta}
\end{equation}
Using Eq. \ref{SIeq:drexhage_gamma} we find that:
\begin{equation}
\frac{\gamma_{\mathrm{tot}}(x)}{\gamma_{\mathrm{tot}, 0}} = 1 + \eta_0 \left[\frac{3\sin x}{2x^3} - \frac{3\cos x}{2x^2} - \frac{3\sin x}{2x} \right]
\label{SIeq:drexhage_gamma_eta_final}
\end{equation}

For a 2D dipole the case is different.
Assuming that the dipole has perfect transverse coherence and thus emits only into forward and backward plane wave modes, the modification of $\gamma_r$ is determined by the interference between  $-e^{ik_0d _\mathrm{opl}}$ and $1$.
Here $k = \frac{2\pi}{\lambda_0}$ is the wavenumber of the light in vacuum and $d_\mathrm{opl}$ is the total {\em optical} path length traversed by the backwards-emitted wave until it comes back to the 2D dipole.
The negative sign comes from the phase flip on reflection.
Thus, we can write down the modification of radiative decay rate:
\begin{equation}
\frac{\gamma_r}{\gamma_{r,0}} = \operatorname{Re}\left[1 -e^{ik_0d_\mathrm{opl}} \right] = 1 - \cos (k_0 d_\mathrm{opl})
\label{SIeq:2D_gamma}
\end{equation}
Identifying that $d_\mathrm{opl} = 2z$:
\begin{equation}
\frac{\gamma_r}{\gamma_{r,0}} = 1 - \cos(2k_0 z) = 1 - \cos(x)
\label{SIeq:2D_gamma_final}
\end{equation}
Using Eq. \ref{SIeq:drexhage_gamma_eta}, which holds for a 2D dipole as well, we find that:
\begin{equation}
\frac{\gamma_{\mathrm{tot}}(x)}{\gamma_{\mathrm{tot},0}} = 1 - \eta_0 \cos(x)
\label{SIeq:2D_gamma_eta}
\end{equation}
\newpage

